\documentclass[pra,twocolumn,showpacs]{revtex4}

\usepackage{amsfonts,amssymb,amsmath,amsthm}

\newtheorem{theorem}{Theorem}

\newtheorem{lemma}{Lemma}
\newtheorem{dfn}{Definition}

\newcommand{\ZZ}{\mathbb{Z}}
\newcommand{\CC}{\mathbb{C}}

\newcommand{\FF}{\mathbb{F}}
\newcommand{\calL}{{\cal L }}
\newcommand{\calH}{{\cal H }}
\newcommand{\calG}{{\cal G }}

\newcommand{\calO}{{\cal O }}
\newcommand{\la}{\langle}
\newcommand{\ra}{\rangle}

\newcommand*{\Cl}[1]{{\cal C}_{#1}}
\newcommand{\bydef}{\stackrel{\mathrm{def}}{=}}

\newcommand{\ep}{\epsilon}
\newcommand{\sx}{\sigma^x}
\newcommand{\sy}{\sigma^y}
\newcommand{\sz}{\sigma^z}

\newcommand{\nn}{\nonumber}
\newcommand{\ba}{\begin{array}}
\newcommand{\ea}{\end{array}}
\newcommand{\sm}{\sigma}

\newcommand{\Tt}{$T$}
\newcommand{\Hh}{$H$}

\usepackage{graphicx}% Include figure files
\usepackage{dcolumn}% Align table columns on decimal point
\usepackage{bm}% bold math

\DeclareMathOperator*{\tr}{\mathop{\mathrm{Tr}}}
\DeclareMathOperator*{\CSS}{\mathop{\mathrm{CSS}}}

\begin{document}

\title{Universal quantum computation with ideal Clifford gates
and noisy ancillas}

\author{Sergey Bravyi}
\email[E-mail:~]{serg@cs.caltech.edu}
\author{Alexei Kitaev}
\email[E-mail:~]{kitaev@iqi.caltech.edu}
\affiliation{Institute for Quantum Information,
California Institute of Technology,\\
Pasadena, 91125 CA, USA.}

\date{\today}

\begin{abstract}
We consider a model of quantum computation in which the set of elementary
operations is limited to Clifford unitaries, the creation of the state
$|0\rangle$, and qubit measurement in the computational basis. In addition, we
allow the creation of a one-qubit ancilla in a mixed state $\rho$, which
should be regarded as a parameter of the model. Our goal is to determine for
which $\rho$ universal quantum computation (UQC) can be efficiently
simulated. To answer this question, we construct purification protocols that
consume several copies of $\rho$ and produce a single output qubit with higher
polarization. The protocols allow one to increase the polarization only along
certain ``magic'' directions. If the polarization of $\rho$ along a magic
direction exceeds a threshold value (about 65\%), the purification
asymptotically yields a pure state, which we call a magic state.  We show that
the Clifford group operations combined with magic states preparation are
sufficient for UQC. The connection of our results with the Gottesman-Knill
theorem is discussed.
\end{abstract}


\maketitle

%%%%%%%%%%%%%%%%%%%%%%%%%%%%%%%%%%%%%%%%%%%%%%%%%%%%%%%%%%%%%%%
%%%%%%%%%%%%%%%%%%%%%%%%%%%%%%%%%%%%%%%%%%%%%%%%%%%%%%%%%%%%%%%
%%%%%%%%%%%%%%%%%%%%%%%%%%%%%%%%%%%%%%%%%%%%%%%%%%%%%%%%%%%%%%%
%%%%%%%%%%%%%%%%%%%%%%%%%%%%%%%%%%%%%%%%%%%%%%%%%%%%%%%%%%%%%%%
%%%%%%%%%%%%%%%%%%%%%%%%%%%%%%%%%%%%%%%%%%%%%%%%%%%%%%%%%%%%%%%
\section{\label{sec:intro} Introduction and summary}

The theory of fault-tolerant quantum computation defines an important number
called the error threshold. If the physical error rate is less than the
threshold value $\delta$, it is possible to stabilize computation by
transforming the quantum circuit into a fault-tolerant form where errors can
be detected and eliminated. However, if the error rate is above the threshold,
then errors begin to accumulate, which results in rapid decoherence and
renders the output of the computation useless. The actual value of $\delta$
depends on the error correction scheme and the error model.
Unfortunately, this number seems to be
rather small for all known schemes. Estimates vary from $10^{-6}$
(see~\cite{KLZ97}) to $10^{-4}$ (see~\cite{Zalka96,Steane97,DKLP01}), which is
hardly achievable with the present technology.

In principle, one can envision a situation in which qubits do not decohere,
and a subset of the elementary gates is realized {\it exactly} due to special
properties of the physical system. This scenario could be realized
experimentally using spin, electron, or other many-body systems with
topologically ordered ground states. Excitations in two-dimensional
topologically ordered systems are anyons --- quasiparticles with unusual
statistics described by non-trivial representations of the braid group. If we
have sufficient control of anyons i.e., are able to move them around each
other, fuse them, and distinguish between different particle types, then we
can realize some set of unitary operators and measurements exactly. This set
may or may not be computationally universal. While the universality can be
achieved with sufficiently nontrivial types of
anyons~\cite{Mochon04,Kitaev97,FLW00,FKLW01}, more realistic systems offer only
decoherence protection and an incomplete set of topological gates.
(See~\cite{MR91,NW96} about non-Abelian anyons in quantum Hall systems
and~\cite{DV01,FI02} about topological orders in Josephson junction arrays.)
Nevertheless, universal computation is possible if we introduce some
additional operations (e.g., measurements by Aharonov-Bohm
interference~\cite{Preskill97} or some gates that are not related to topology
at all). Of course, these non-topological operations cannot be implemented
exactly and thus are prone to errors.

In this situation, the threshold error rate $\delta$ may become significantly
larger than the values given above because we need to correct only errors of
certain special type and we introduce a smaller amount of error in the
correction stage. The main purpose of the present paper is to illustrate this
statement by a particular computational model.

The model  is built upon the {\it Clifford group} --- the
group of unitary operators that map the group of Pauli operators to itself under conjugation.
The set of elementary operations is divided into two parts:
$\calO=\calO_{ideal}\cup\calO_{faulty}$.  Operations from $\calO_{ideal}$ are
assumed to be perfect.    
 We list these operations below:
\begin{itemize}
\item Prepare a qubit in the state $|0\ra$;
\item Apply unitary operators from the Clifford group;
\item Measure an eigenvalue of a Pauli operator ($\sx$, $\sy$, or $\sz$) on
any qubit.
\end{itemize}
Here we mean non-destructive projective measurement. We also assume that no
errors occur between the operations.

It is well-known that these operations are not sufficient for universal
quantum computation (unless a quantum computer can be efficiently simulated on
a classical computer). More specifically, the Gottesman-Knill theorem states
that by operations from $\calO_{ideal}$ one can only obtain quantum states of
a very special form called {\it stabilizer states}. Such a state can be
specified as an intersection of eigenspaces of pairwise commuting Pauli
operators, which are referred to as {\it stabilizers}.  Using the stabilizer
formalism, one can easily simulate the evolution of the state and the
statistics of measurements on a classical probabilistic computer
(see~\cite{G97} or a textbook~\cite{CN} for more details).

The set $\calO_{faulty}$ describes faulty operations. In our model, it
consists of just one operation:
\begin{itemize}
\item Prepare an ancillary qubit in a mixed state $\rho$.
\end{itemize}
The state $\rho$ should be regarded as a parameter of the model.  From the
physical point of view, $\rho$ is mixed due to imperfections of the
preparation procedure (entanglement of the ancilla with the environment,
thermal fluctuations, etc.). An essential requirement is that by preparing $n$
qubits we obtain the state $\rho^{\otimes n}$ i.e., all ancillary qubits are
independent. The independence assumption is similar to the uncorrelated errors
model in the standard fault-tolerant computation theory.

Our motivation for including 
all Clifford group gates into $\calO_{ideal}$ relies mostly
on the recent progress in the fault-tolerant implementation of such
gates.
For instance, using a concatenated stabilizer code with good error
correcting properties to encode each qubit
and applying gates transversally (so that errors do not propogate
inside code blocks) one can implement Clifford gates with 
an arbitrary high precision, see~\cite{G97a}. However these
nearly perfect gates act on  {\it encoded} qubits.
To establish a correspondence with our model, one needs to
prepare an {\it encoded} ancilla in the state $\rho$. 
It can be done using the schemes
for fault-tolerant encoding of an arbitrary {\it known} one-qubit state
described by Knill in~\cite{Knill04_feb}. 
In the more recent paper~\cite{Knill04_april} Knill 
constructed a novel scheme of fault-tolerant quantum computation
which combines
(i) the teleported computing and error correction technique 
by Gottesman and Chuang~\cite{Nature99};
(ii) the method of purification of CSS states by D\"ur and Briegel~\cite{DurBriegel03};
and (iii) the magic states distillation algorithms described in the present paper. 
As was argued in~\cite{Knill04_april}, this scheme is likely to yield much higher
value for the threshold $\delta$ (may be up to $1\%$).

Unfortunately, ideal implementation of the Clifford group can not be currently
achieved in any realistic physical system with a topological order.
What universality classes of anyons  allow one to implement all Clifford
group gates (but do not allow one to simulate UQC) is an interesting open
problem. 

To fully utilize the potential of our model, we allow {\it adaptive}
computation. It means that a description of 
an operation to be performed at
the step $t$ may be a function of all 
measurement outcomes at the steps $1,\ldots,t-1$.
(For even greater
generality, the dependence may be probabilistic. This assumption does not
actually strengthen the model since tossing a fair coin can be simulated using
$\calO_{ideal}$.) At this point, we need to be careful because the proper
choice of  operations should not only be defined mathematically --- it
should be computed by some {\it efficient algorithm}.  In all protocols
described below, the algorithms will actually be very simple. (Let us 
point out that dropping the computational complexity restriction still 
leaves a non-trivial problem: can we prepare an arbitrary multiqubit pure state with
any given fidelity using only operations from the basis $\calO$?) 

The main question that we address in this paper is as follows: For which
density matrices $\rho$ can one efficiently simulate universal quantum
computation by adaptive computation in the basis $\calO$?
 
It will be convenient to use the Bloch sphere representation of one-qubit
states:
\[
\rho=\frac12 \left( I + \rho_x \sx + \rho_y \sy + \rho_z \sz \right).
\]
The vector $(\rho_x,\rho_y,\rho_z)$ will be referred to as the {\it
polarization vector} of $\rho$. Let us first consider the subset of states
satisfying
\[
\label{inside_T}
|\rho_x| + |\rho_y| + |\rho_z| \le 1.
\]
This inequality says that the vector $(\rho_x,\rho_y,\rho_z)$ lies inside the
octahedron $O$ with vertices $(\pm 1,0,0)$, $(0,\pm 1,0)$, $(0,0,\pm 1)$, see
Fig.~1. The six vertices of $O$ represent the six eigenstates of the Pauli
operators $\sx$, $\sy$, and $\sz$. We can prepare these states by operations
from $O_{ideal}$ only. Since $\rho$ is a convex linear combination
(probabilistic mixture) of these states, we can prepare $\rho$ by operations
from $\calO_{ideal}$ and by tossing a coin with suitable weights. Thus we can
rephrase the Gottesman-Knill theorem in the following way.
\begin{theorem}
\label{Gottesman-Knill}
Suppose the polarization vector $(\rho_x,\rho_y,\rho_z)$ of the state $\rho$
belongs to the convex hull of $(\pm 1,0,0)$, $(0,\pm 1,0)$, $(0,0,\pm 1)$.
Then any adaptive computation in the basis $\calO$ can be efficiently
simulated on a classical probabilistic computer.
\end{theorem}

\begin{figure}
\includegraphics[scale=0.4]{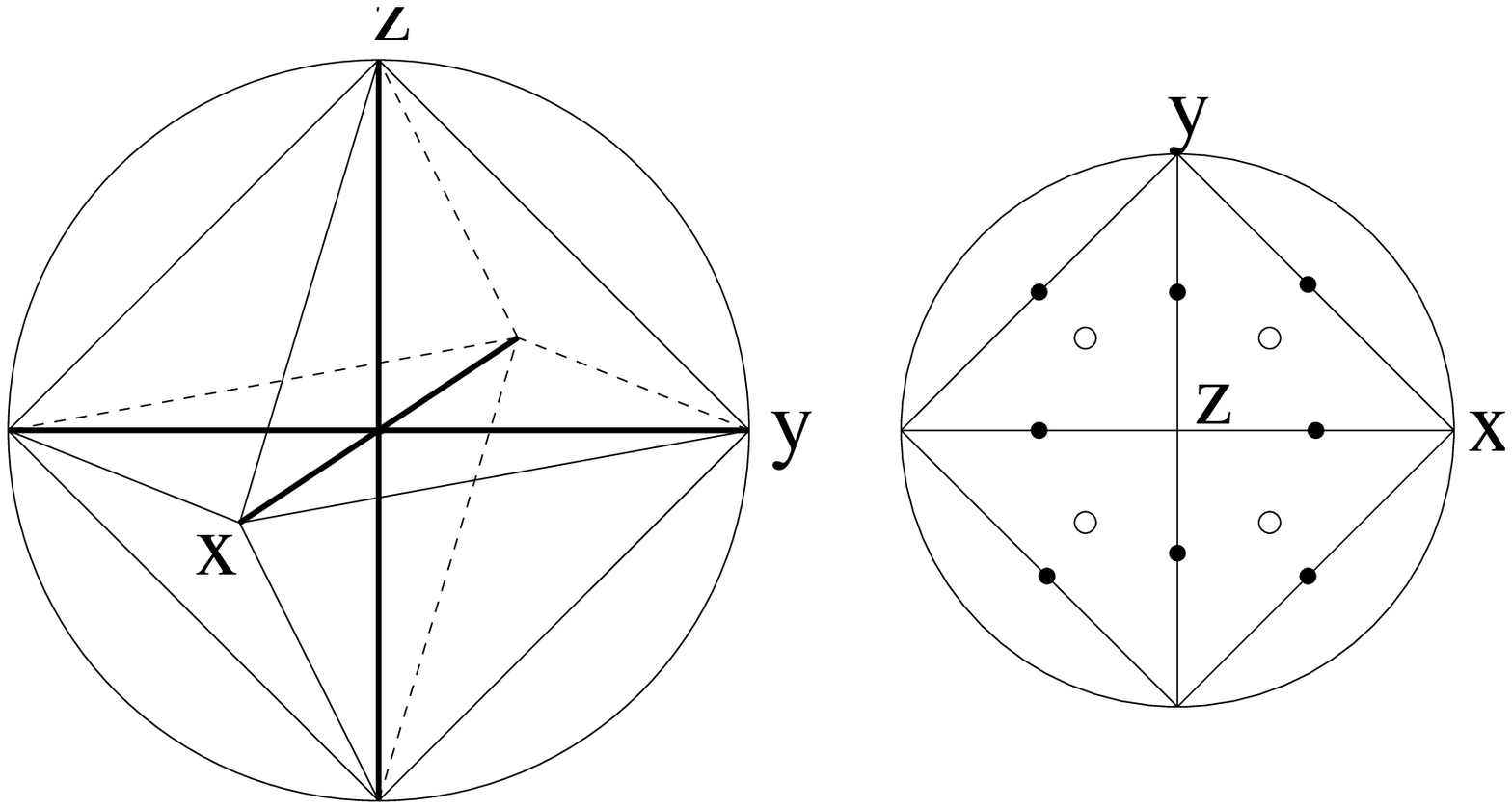}
\caption{On the left: The Bloch sphere and the octahedron $O$.  On the right:
The octahedron $O$ projected on the $x$-$y$ plane.  The magic states
correspond to the intersections of the symmetry axes of $O$ with the Bloch
sphere.  The empty and filled circles represent \Tt-type and \Hh-type magic
states, respectively.}
\end{figure}

This observation leads naturally to the following question: is it true that
UQC can be efficiently simulated whenever $\rho$ lies in the exterior of the
octahedron $O$? In an attempt to provide at least a partial answer, we prove
the universality for a large set of states.  Specifically, we
construct two particular schemes of UQC simulation based on a method which we
call {\it magic states distillation}. Let us start by defining the magic
states.  
\begin{dfn}
Consider  pure states $|H\ra, |T\ra \in \CC^2$ such that
\[
|T\ra\la T| = \frac12\left[ I + 
\frac1{\sqrt{3}} \left( \sx + \sy + \sz \right)\right]
\]
and
\[
|H\ra\la H| = \frac12 \left[ I + 
\frac1{\sqrt{2}} \left( \sx + \sz \right)\right].
\]
The images of $|T\ra$ and $|H\ra$ under the action of one-qubit Clifford
operators are called magic states of \Tt-type and \Hh-type, respectively.
\end{dfn}

\noindent
(This notation is chosen since $|H\ra$ and $|T\ra$ are eigenvectors of certain
Clifford group operators: the Hadamard gate $H$ and the operator usually
denoted $T$, see Eq.~(\ref{T-op}).) 
Denote the one-qubit Clifford group by $\Cl{1}$.
Overall, there are eight magic states of
\Tt-type, $\{ U|T\ra, \; U\in \Cl{1}\}$ (up to a phase) and twelve states of
\Hh-type, $\{ U|H\ra,\; U\in \Cl{1}\}$, see Fig.~1.
Clearly, the polarization
vectors of magic states are in one-to-one correspondence with rotational
symmetry axes of the octahedron $O$ (\Hh-type states correspond to $180^\circ$
rotations and \Tt-type states correspond to $120^\circ$ rotations). The role of
magic states in our construction is two-fold.  First, adaptive computation in
the basis $\calO_{ideal}$ together with the preparation of magic states (of
either type) allows one to simulate UQC. (See Sec.~\ref{sec:UQC}.) Second, by
adaptive computation in the basis $\calO_{ideal}$ one can ``purify'' imperfect
magic states. It is a rather surprising coincidence that one and the same
state can comprise both of these properties, and that is the reason why we call  
them magic states.

More exactly, a magic state distillation procedure yields one copy of a magic
state (with any desired fidelity) from several copies of the state $\rho$,
provided that the initial fidelity between $\rho$ and the magic state to be
distilled is large enough. In the course of distillation, we use only
operations from the set $\calO_{ideal}$.  By constructing two particular
distillation schemes, for \Tt-type and \Hh-type magic states, respectively, we
prove the following theorems.

\begin{theorem}
\label{T-theorem}
Let $F_T(\rho)$ be the maximum fidelity between $\rho$ and a \Tt-type magic
state i.e.,
\[
F_T(\rho)=\max_{U\in \Cl{1}} \sqrt{\la T|U^\dag \rho U |T\ra}.
\]
Adaptive computation in the basis $\calO=\calO_{ideal}\cup\{\rho\}$ allows one
to simulate universal quantum computation whenever
\[
F_T(\rho)>F_T\bydef \left[\frac12\left( 1+\sqrt{\frac37}\right)\right]^\frac12
 \approx 0.910.
\]
\end{theorem}

\begin{theorem}
\label{H-theorem}
Let $F_H(\rho)$ be the maximum fidelity between $\rho$ and an \Hh-type magic
state,
\[
F_H(\rho)=\max_{U\in \Cl{1}} \sqrt{\la H|U^\dag \rho U |H\ra}.
\]
Adaptive computation in the basis $\calO=\calO_{ideal}\cup\{\rho\}$ allows one
to simulate universal quantum computation whenever
\[
F_H(\rho)>F_H\approx 0.927.
\]
\end{theorem}

The quantities $F_T$ and $F_H$ have the meaning of threshold fidelity since
our distillation schemes increase the polarization of $\rho$, converging to a
magic state as long as the inequalities $F_T(\rho)>F_T$ or $F_H(\rho)>F_H$ are
fulfilled. If they are not fulfilled, the process converges to the maximally
mixed state. The conditions stated in the theorems can also be understood in
terms of the polarization vector $(\rho_x,\rho_y,\rho_z)$.  Indeed, let us
associate a ``magic direction'' with each of the magic states. Then
Theorems~\ref{T-theorem}, \ref{H-theorem} say that the distillation is
possible if there is a \Tt-direction such that the projection of the vector
$(\rho_x,\rho_y,\rho_z)$ onto that \Tt-direction exceeds 
the threshold value of $2F_{T}^{2}-1\approx 0.655$,
or if the projection on some of the $H$-directions is greater than
$2F_{H}^{2}-1\approx 0.718$.

Let us remark that, although the proposed distillation schemes are probably
not optimal, the threshold fidelities $F_T$ and $F_H$ can not be improved
significantly. Indeed, it is easy to check that the octahedron $O$
corresponding to probabilistic mixtures of stabilizer states can be defined as
\[
\calO = \{ \rho \; : \;  F_T(\rho)\le F_T^*\},
\]
where 
\[
F_T^* \bydef \left[\frac12\left( 1+\sqrt{\frac13}\right)\right]^\frac12
\approx 0.888.
\]
It means that $F_T^*$ is a lower bound on the threshold fidelity $F_T$ for any
protocol distilling \Tt-type magic states. Thus any potential improvement to
Theorem~\ref{T-theorem} may only decrease $F_T$ from $0.910$ down to
$F_T^*=0.888$. From a practical perspective, the difference between these two
numbers is not important.

On the other hand, such an improvement would be of great theoretical
interest. Indeed, if Theorem~\ref{T-theorem} with $F_T$ replaced by $F_T^*$ is
true, it would imply that the Gottesman-Knill theorem provides necessary and
sufficient conditions for the classical simulation, and that a transition from
classical to universal quantum behavior occurs at the boundary of the
octahedron $O$.  This kind of transition has been discussed in context of a
general error model~\cite{Aharonov99}. Our model is simpler, which gives hope
for sharper results.

By the same argument, one can show that the quantity
\[
F_H^* \bydef \max_{\rho \in O} \sqrt{\la H|\rho |H\ra} =
\left[\frac12\left( 1+\sqrt{\frac12}\right)\right]^\frac12
\approx 0.924.
\]
is a lower bound on the threshold fidelity $F_H$ for any protocol
distilling \Hh-type magic states.

A similar approach to UQC simulation was suggested in the
work~\cite{Dennis01}, where Clifford group operations were used to distill the
entangled three-qubit state $|000\ra + |001\ra + |010\ra + |100\ra$, which is
necessary for the realization of the Toffoli gate.

The rest of the paper is organized as follows.  Section~\ref{sec:Clifford}
contains some well-known facts about the Clifford group and stabilizer
formalism, which will be used throughout the paper.  In Sec.~\ref{sec:UQC}
we prove that magic states together with operations from $\calO_{ideal}$ are
sufficient for UQC. In Sec.~\ref{sec:connection} ideal magic are substituted by
faulty ones and the error rate that our simulation 
algorithm can tolerate is estimated.
In Sec.~\ref{sec:T} we describe a distillation
protocol for \Tt-type magic states. This protocol is based on the well-known
five-qubit quantum code. In Sec.~\ref{sec:H} a distillation protocol for
\Hh-type magic states is constructed. It is based on a certain CSS stabilizer
code that encodes one qubit into 15 and admits a nontrivial
automorphism~\cite{KLZ96}. Specifically, the bitwise application of a certain
{\it non-Clifford} unitary operator preserves the code subspace and effects
the same operator on the encoded qubit. We conclude with a brief summary and a
discussion of open problems.

%%%%%%%%%%%%%%%%%%%%%%%%%%%%%%%%%%%%%%%%%%%%%%%%%%%%%%%%%%%%
%%%%%%%%%%%%%%%%%%%%%%%%%%%%%%%%%%%%%%%%%%%%%%%%%%%%%%%%%%%%
%%%%%%%%%%%%%%%%%%%%%%%%%%%%%%%%%%%%%%%%%%%%%%%%%%%%%%%%%%%%
%%%%%%%%%%%%%%%%%%%%%%%%%%%%%%%%%%%%%%%%%%%%%%%%%%%%%%%%%%%%
%%%%%%%%%%%%%%%%%%%%%%%%%%%%%%%%%%%%%%%%%%%%%%%%%%%%%%%%%%%%
\section{\label{sec:Clifford} The Clifford group, stabilizers, and
syndrome  measurements}

Let $\Cl{n}$ denote the $n$-qubit \emph{Clifford group}. Recall that it is a
finite subgroup of $U(2^n)$ generated by the Hadamard gate $H$ (applied to any qubit),
the phase-shift gate $K$ (applied to any qubit), and the controlled-not gate
$\Lambda(\sx)$ (which may be applied to any pair qubits).
\begin{equation}\label{gates} H= \frac1{\sqrt{2}}\left(
\ba{cc} 1& 1 \\ 1 & -1 \\ \ea \right),\, K=\left( \ba{cc} 1 & 0 \\ 0 & i \\
\ea \right),\, \Lambda(\sx)=\left( \ba{cc} I & 0 \\ 0 & \sx \\ \ea \right).
\end{equation}
The Pauli operators $\sx$, $\sy$, $\sz$ belong to $\Cl{1}$, for instance,
$\sz=K^2$ and $\sx=H K^2 H$. The \emph{Pauli group} $P(n)\subset \Cl{n}$ is
generated by the Pauli operators acting on $n$ qubits. It is
known~\cite{orthogeom} that the Clifford group $\Cl{n}$ augmented by scalar
unitary operators $e^{i\varphi} I$ coincides with the normalizer of $P(n)$ in
the unitary group $U(2^n)$.  Hermitian elements of the Pauli group are of
particular importance for quantum error correction theory; they are referred
to as {\it stabilizers}. These are operators of the form
\[
\pm \sigma^{\alpha_{1}}\otimes\cdots\otimes\sigma^{\alpha_{n}},
\quad\ \alpha_{j}\in\{0,x,y,z\},
\]
where $\sigma^{0}=I$.  Let us denote by $S(n)$ the set of all $n$-qubit
stabilizers:
\[
S(n)=\{ S\in P(n) \; : \; S^\dag=S\}.
\]
For any two stabilizers $S_1, S_2$ we have $S_1 S_2 = \pm S_2 S_1$ and
$S_1^2=S_2^2=I$. It is known that for any set of pairwise commuting
stabilizers $S_1,\ldots,S_k \in S(n)$ there exists a unitary operator $V\in
\Cl{n}$ such that
\[
V S_j V^\dag = \sz[j], \quad j=1,\ldots, k,
\]
where $\sz[j]$ denotes the operator $\sz$ applied to the $j$-th qubit,
e.g., $\sz[1]=\sz\otimes I \otimes \cdots \otimes I$. 

These properties of the Clifford group allow us to introduce a very useful
computational procedure which can be realized by operations from
$\calO_{ideal}$. Specifically, we can perform a joint non-destructive
eigenvalue measurement for any set of pairwise commuting stabilizers
$S_1,\ldots, S_k \in S(n)$.  The outcome of such a measurement is a sequence
of eigenvalues $\lambda=(\lambda_1,\ldots, \lambda_k)$,\, $\lambda_j = \pm 1$,
which is usually called a {\it syndrome}. For any given outcome, the quantum
state is acted upon by the projector
\[
\Pi_\lambda = \prod_{j=1}^k \frac12 \bigl(I + \lambda_j S_j \bigr).
\]

Now, let us consider a computation that begins with an arbitrary state and
consists of operations from $\calO_{ideal}$. It is clear that we can defer all
Clifford operations until the very end if we replace the Pauli measurements by
general syndrome measurements. Thus the most general transformation that can
be realized by $\calO_{ideal}$ is an {\it adaptive syndrome measurement},
meaning that the choice of the stabilizer $S_j$ to be measured next depends on
the previously measured values of $\lambda_1,\ldots,\lambda_{j-1}$. In
general, this dependence may involve coin tossing. Without loss of generality
one can assume that $S_j$ commutes with all previously measured stabilizers
$S_1,\ldots,S_{j-1}$ (for all possible values of
$\lambda_1,\ldots,\lambda_{j-1}$ and coin tossing outcomes).  Adaptive
syndrome measurement has been used in the work~\cite{AG03} to distill
entangled states of a bipartite system by local operations.

%%%%%%%%%%%%%%%%%%%%%%%%%%%%%%%%%%%%%%%%%%%%%%%%%%%%%%%%%%%%%%%%%%%%%%%
%%%%%%%%%%%%%%%%%%%%%%%%%%%%%%%%%%%%%%%%%%%%%%%%%%%%%%%%%%%%%%%%%%%%%%%
%%%%%%%%%%%%%%%%%%%%%%%%%%%%%%%%%%%%%%%%%%%%%%%%%%%%%%%%%%%%%%%%%%%%%%%
%%%%%%%%%%%%%%%%%%%%%%%%%%%%%%%%%%%%%%%%%%%%%%%%%%%%%%%%%%%%%%%%%%%%%%%
%%%%%%%%%%%%%%%%%%%%%%%%%%%%%%%%%%%%%%%%%%%%%%%%%%%%%%%%%%%%%%%%%%%%%%%
\section{\label{sec:UQC} Universal quantum computation with magic states}

In this section, we show that operations from $\calO_{ideal}$ are sufficient
for universal quantum computation if a supply of {\it ideal } magic states is also
available. First, consider a one-qubit state
\begin{equation}
\label{theta}
|A_\theta\ra=2^{-1/2} (|0\ra + e^{i\theta} |1\ra)
\end{equation}
and suppose that $\theta$ is not a multiple of $\pi/2$.
We now describe a procedure that implements the phase shift gate
\[
\Lambda(e^{i\theta}) = \left( \ba{cc} 1 & 0 \\ 0 & e^{i\theta} \\ \ea \right)
\]
by consuming several copies of $|A_\theta\ra$ and using only operations from
$\calO_{ideal}$.

Let $|\psi\ra = a |0\ra + b |1\ra$ be the unknown initial state which should
be acted on by $\Lambda(e^{i\theta})$. Prepare the state
$|\Psi_0\ra=|\psi\ra\otimes|A_\theta\ra$ and measure the stabilizer
$S_1=\sm^z\otimes \sm^z$. Note that both outcomes of this measurement appear
with probability $1/2$. If the outcome is `$+1$', we are left with the state
\[
|\Psi_1^{+}\ra=\bigl(a |0,0\ra + b e^{i\theta} |1,1\ra\bigr).
\]
In the case of `$-1$' outcome, the resulting state is
\[
|\Psi_1^{-}\ra=\bigl(a e^{i\theta} |0,1\ra + b |1,0\ra\bigr).
\]
Let us XOR the first qubit into the second qubit (i.e., apply the operator
$\Lambda(\sx)$).  The above two states are mapped to
\[
\begin{array}{l}
|\Psi_2^{+}\ra= {\rm XOR}[1,2]|\Psi_1^{+}\ra =
\bigl(a |0\ra + b e^{i\theta} |1\ra\bigr)\otimes|0\ra,
\\[5pt]
|\Psi_2^{-}\ra= {\rm XOR}[1,2]|\Psi_1^{-}\ra =
\bigl(a e^{i\theta} |0\ra + b |1\ra\bigr)\otimes|1\ra.
\end{array}
\]
Now the second qubit can be discarded, and we are left with the state $a|0\ra
+ b e^{\pm i\theta}|1\ra$, depending upon the measured eigenvalue.  Thus the
net effect of this circuit is the application of a unitary operator that is
chosen randomly between $\Lambda(e^{i\theta})$ or $\Lambda(e^{-i\theta})$ (and
we know which of the two possibilities has occurred).

Applying the circuit repeatedly, we effect the transformations
$\Lambda(e^{ip_1\theta}),\,\Lambda(e^{ip_2\theta}),\ldots$ for some integers
$p_1,p_2,\ldots$ which obey the random walk statistics. It is well known that
such a random walk visits each integer with the probability one. It means that
sooner or later we will get $p_k=1$ and thus realize the desired operator
$\Lambda(e^{i\theta})$.  The probability that we will need more than $N$ steps
to succeed can be estimated as $c N^{-1/2}$ for some constant $c>0$.  Note
also that if $\theta$ is a rational multiple of $2\pi$, we actually have a
random walk on a cyclic group $\ZZ_q$.  In this case, the probability that we
will need more than $N$ steps decreases exponentially with $N$.

The magic state $|H\ra$ can be explicitly written in the standard basis as
\begin{equation}
\label{H-state}
|H\ra=\cos{\left(\frac{\pi}8\right)} |0\ra
+ \sin{\left(\frac{\pi}8\right)} |1\ra.
\end{equation}
Note that $HK|H\ra = e^{i\pi/8}|A_{-\pi/4}\ra$. So if we are able to prepare
the state $|H\ra$, we can realize the operator $\Lambda(e^{-i\pi/4})$.  It
does not belong to the Clifford group. Moreover, the subgroup of $U(2)$
generated by $\Lambda(e^{-i\pi/4})$ and $\Cl{1}$ is dense in $U(2)$,
see~\footnote{Recall that the action of the Clifford group $\Cl{1}$ on the set
of operators $\pm\sm^x$, $\pm\sm^y$, $\pm\sm^z$ coincides with the action of
rotational symmetry group of a cube on the set of unit vectors $\pm e_x$,
$\pm e_y$, $\pm e_z$, respectively.}. Thus the operators from $\Cl{1}$ and
$\Cl{2}$ together with $\Lambda(e^{-i\pi/4})$ constitute a universal basis
for quantum computation.

The magic state $|T\ra$ can be explicitly written in the standard basis:
\begin{equation}
\label{T-state}
|T\ra = \cos\beta\, |0\ra + e^{i\frac{\pi}4} \sin\beta\, |1\ra, \quad
\cos{(2\beta)}=\frac1{\sqrt{3}}.
\end{equation}
Let us prepare an initial state $|\Psi_0\ra=|T\ra\otimes|T\ra$ and measure the
stabilizer $S_1=\sm^z\otimes\sm^z$.  The outcome `$+1$' appears with probability
$p_+=\cos^{4}\beta + \sin^{4}\beta = 2/3$.  If the outcome is `$-1$', we discard
the reduced state and try again, using a fresh pair of magic states.  (On 
average, we need three copies of the $|T\ra$ state to get the outcome `$+1$'.)
The reduced state corresponding to the outcome `$+1$' is
\[
|\Psi_1\ra=\cos{\gamma}\,|0,0\ra + i\sin{\gamma}\,|1,1\ra, \quad
\gamma=\frac{\pi}{12}.
\]
Let us XOR the first qubit into the second and discard the second qubit.
We arrive at the state
\[
|\Psi_2\ra=\cos{\gamma}\,|0\ra + i\sin{\gamma}\,|1\ra.
\]
Next apply the Hadamard gate $H$:
\[
|\Psi_3\ra = H|\Psi_2\ra =
2^{-1/2} e^{i\gamma} \bigl(|0\ra + e^{-2i\gamma}|1\ra\bigr)=|A_{-\pi/6}\ra.
\]
We can use this state as described above to realize the operator
$\Lambda(e^{-i\pi/6})$.  It is easy to check that Clifford operators together
with $\Lambda(e^{-i\pi/6})$ constitute a universal set of unitary gates.

Thus we have proved that the sets of operations
$\calO_{ideal}\cup\{|H\rangle\}$ and $\calO_{ideal}\cup\{|T\rangle\}$ are
sufficient for universal quantum computation.

%%%%%%%%%%%%%%%%%%%%%%%%%%%%%%%%%%%%%%%%%%%%%%%%%%%%%%%%%%%%%%%%%%
%%%%%%%%%%%%%%%%%%%%%%%%%%%%%%%%%%%%%%%%%%%%%%%%%%%%%%%%%%%%%%%%%%
%%%%%%%%%%%%%%%%%%%%%%%%%%%%%%%%%%%%%%%%%%%%%%%%%%%%%%%%%%%%%%%%%%
%%%%%%%%%%%%%%%%%%%%%%%%%%%%%%%%%%%%%%%%%%%%%%%%%%%%%%%%%%%%%%%%%%
%%%%%%%%%%%%%%%%%%%%%%%%%%%%%%%%%%%%%%%%%%%%%%%%%%%%%%%%%%%%%%%%%%
\section{\label{sec:connection} Error analysis}

To establish a connection between the simulation algorithms described in Sec.~\ref{sec:UQC}
and the universality  theorems  stated in the introduction we have to
substitute {\it ideal} magic states by {\it faulty} ones.
Before doing that let us discuss  the ideal case in more details.
Suppose that a quantum circuit to be simulated uses a gate basis in
which the only non-Clifford gate is the phase shift $\Lambda(e^{-i\pi/4})$
or $\Lambda(e^{-i\pi/6})$.
One can apply the algorithm of Sec.~\ref{sec:UQC} to simulate each
non-Clifford gate  independently. To avoid fluctuations in
the number of magic states consumed at each round, let us set
a limit of $K$ magic states per round,
where $K$ is a parameter to be chosen later.
As was pointed out in Sec.~\ref{sec:UQC},
the probability for some particular simulation round to ``run out of budget'' 
scales as $\exp{(-\alpha K)}$ for some constant $\alpha>0$. 
If at least one simulation round runs out of budget, we declare a failure and
the whole simulation must be aborted. 
Denote the total number of non-Clifford gates  in the circuit by $L$.
The probability $p_a$ for the whole  simulation to be aborted can be estimated as 
\[
p_a \sim 1- (1-\exp{(-\alpha K)})^L \sim L\exp{(-\alpha K)}\ll 1,
\]
provided that $L\exp{(-\alpha K)}\ll 1$. We will assume 
\[
K\gtrsim \alpha^{-1} \log{L},
\]
so the abort probability can be neglected.

Each time the algorithm requests an ideal magic state, it actually
receives a slightly non-ideal one. Such nearly
perfect magic states must be prepared  using the distillation methods
described in Sec.~\ref{sec:T},\ref{sec:H}. Let us estimate an
affordable error rate $\ep_{out}$ for {\it distilled} magic states. Since there are
$L$ non-Clifford gates in the circuit, one can tolerate 
an error rate of the order $1/L$ in implementation of these gates~\footnote{This
fault-tolerance does not require any redundancy in the implementation of the circuit
(e.g. the use of concatenated codes). It is achived automatically
because in the worst case the error probability accumulates linearly
in the number of gates.
In our model only non-Clifford gates are faulty.}.
Each non-Clifford gate requires $K\sim \log{L}$ magic states.
Thus the whole simulation is reliable enough if one chooses
\begin{equation}\label{e(L)}
\ep_{out}\sim 1/(L\log{L}).
\end{equation}

What are the resources needed to distill one copy of a magic state
with the error rate $\ep_{out}$? To be more specific, let us talk about
\Hh-type states. It will be shown in Sec.~\ref{sec:H} that the number $n$
of raw (undistilled) ancillas  needed to
distill one copy of the $|H\ra$ magic state with an error rate 
not exceeding $\ep_{out}$ scales as
\[
n\sim (\log{(1/\ep_{out})})^\gamma, \quad \gamma=\log_3{15}\approx 2.5,
\]
see Eq.~(\ref{H-eff}). Taking $\ep_{out}$ from Eq.~(\ref{e(L)}), one gets
\[
n\sim (\log{L})^\gamma.
\]
Since the whole simulation requires $KL\sim L\log{L}$ copies of the distilled 
$|H\ra$ state,
we need 
\[
N\sim L(\log{L})^{\gamma+1}
\]
raw ancillas overall. 

Summarizing, the simulation theorems stated in the introduction 
follow from the following results (the last one will be proved later):
\begin{itemize}
\item
The circuits described in Sec.~\ref{sec:UQC}
allow one to simulate UQC with the sets of operations
$\calO_{ideal}\cup\{|H\rangle\}$ and $\calO_{ideal}\cup\{|T\rangle\}$;

\item 
These circuits work reliably enough if the states $|H\ra$ and $|T\ra$
are slightly noisy, provided  that the error rate does not exceed
$\ep_{out}\sim 1/(L\log{L})$;

\item
A magic state having an error rate $\ep_{out}$ can be prepared
from copies of the raw ancillary state $\rho$ 
using the distillation schemes 
provided that $F_T(\rho)>F_T$ or $F_H(\rho)>F_H$.
The distillation requires resources that are polynomial in $\log{L}$.

\end{itemize}

%%%%%%%%%%%%%%%%%%%%%%%%%%%%%%%%%%%%%%%%%%%%%%%%%%%%%%%%%%%%%%%%%%
%%%%%%%%%%%%%%%%%%%%%%%%%%%%%%%%%%%%%%%%%%%%%%%%%%%%%%%%%%%%%%%%%%
%%%%%%%%%%%%%%%%%%%%%%%%%%%%%%%%%%%%%%%%%%%%%%%%%%%%%%%%%%%%%%%%%%
%%%%%%%%%%%%%%%%%%%%%%%%%%%%%%%%%%%%%%%%%%%%%%%%%%%%%%%%%%%%%%%%%%
%%%%%%%%%%%%%%%%%%%%%%%%%%%%%%%%%%%%%%%%%%%%%%%%%%%%%%%%%%%%%%%%%%
\section{\label{sec:T} Distillation of \Tt-type magic states}

Suppose we are given $n$ copies of a state $\rho$, and our goal is to
distill one copy of the magic state $|T\ra$.  The polarization vector of
$\rho$ can be brought into the positive octant of the Bloch space by a
Clifford group operator, so we can assume that
\[
\rho_x, \rho_y, \rho_z \ge 0.
\]
In this case, the fidelity between $\rho$ and $|T\ra$ is the
largest one among all \Tt-type magic states i.e.,
\[
F_T(\rho)=\sqrt{\la T|\rho| T\ra}.
\]
A related quantity,
\[
\ep = 1 - \la T|\rho| T\ra = \frac12 \left[ 1 - \frac1{\sqrt{3}} (
\rho_x + \rho_y + \rho_z )\right],
\]
will be called the {\it initial error probability}.
By definition, $0\le \ep\le 1/2$.

The output of the distillation algorithm will be some one-qubit
mixed state $\rho_{out}$. To quantify the proximity between $\rho_{out}$
and $|T\ra$, let us define a {\it final  error probability}:
\[
\ep_{out}=1-\la T|\rho_{out}|T\ra.
\]
It will be certain function of $n$ and $\ep$. The asymptotic behavior of this
function for $n\to\infty$ reveals the existence of a {\it threshold error
probability},
\[
\ep_0 = \frac12 \left(1-\sqrt{\frac37}\right) \approx 0.173,
\]
such that for $\ep<\ep_0$ the function $\ep_{out}(n,\ep)$ converges to zero.
We will see that for small $\ep$,
\begin{equation}\label{eoutT}
\ep_{out}(n,\ep) \sim \left(5\ep\right)^{\displaystyle n^\xi}, \quad\
\xi=1/\log_2{30} \approx 0.2.
\end{equation}
On the other hand, if $\ep>\ep_0$, the output state converges to the maximally
mixed state i.e., $\lim_{n\to \infty} \ep_{out}(n,\ep) =1/2$.

Before coming to a detailed description of the distillation algorithm, let us
outline the basic ideas involved in its construction.  The algorithm
recursively iterates an elementary distillation subroutine that transforms
five copies of an imperfect magic state into one copy having a smaller error
probability. This elementary subroutine involves a syndrome measurement for
certain commuting stabilizers $S_1, S_2, S_3, S_4 \in S(5)$.  If the measured
syndrome $(\lambda_1,\lambda_2,\lambda_3,\lambda_4)$ is nontrivial
($\lambda_j=-1$ for some $j$), the distillation attempt fails and the reduced
state is discarded.  If the measured syndrome is trivial ($\lambda_j=1$ for
all $j$), the distillation attempt is successful. Applying a decoding
transformation (a certain Clifford operator) to the reduced state, we
transform it to a single-qubit state. This qubit is the output of the
subroutine.

Our construction is similar to concatenated codes used in many fault-tolerant
quantum computation techniques, but it differs from them in two respects.
First, we do not need to {\it correct} errors --- it suffices only to {\it
detect} them. Once an error has been detected, we simply discard the reduced
state, since it does not contain any valuable information. This allows us to
achieve higher threshold error probability. Second, we do not use quantum
codes in the way for which they were originally designed: in our scheme, the
syndrome is measured on a product state.

The state $|T\ra$ is an eigenstate for the unitary operator
\begin{equation}
\label{T-op}
T = e^{i\pi/4} K H = \frac{e^{i\pi/4} }{\sqrt{2}}
\left( \ba{cc} 1 & 1 \\ i & -i \\ \ea \right) \in \Cl{1}.
\end{equation}
Note that $T$ acts on the Pauli operators as follows~\footnote{The
operator denoted by $T$ in the paper~\cite{G97a} does not coincide
with our $T$. They are related by the substitution $T\to e^{-i\pi/4}T^\dag$
though.}:
\begin{equation}\label{T-action}
T\sx T^\dag = \sz,\quad T\sz T^\dag = \sy, \quad 
T\sy T^\dag =\sx.
\end{equation}
We will denote its eigenstates by $|T_0\ra$ and $|T_1\ra$, so that
\[
T|T_0\ra = e^{+ i\pi/3} |T_0\ra, \quad 
T|T_1\ra = e^{- i\pi/3} |T_1\ra,
\]
\[
|T_{0,1}\ra\la T_{0,1}| = \frac12  \left[ I \pm
\frac1{\sqrt{3}} (\sx + \sy + \sz) \right].
\]
Note that $|T_0\ra\bydef |T\ra$ and $|T_1\ra=\sy H |T_0\ra$ are
\Tt-type magic states. 

Let us apply a dephasing transformation, 
\begin{equation}
\label{diag}
D(\eta)=\frac13 (\eta + T\eta T^\dag + T^\dag \eta T)
\end{equation}
to each copy of the state $\rho$. The transformation $D$ can be realized by
applying one of the operators $I$, $T$, $T^{-1}$ chosen with probability $1/3$
each.  Since
\[
D\bigl(|T_0\ra\la T_1|\bigr)=D\bigl(|T_1\ra\la T_0|\bigr)=0,
\]
we have
\begin{equation}
\label{diag1}
D(\rho)=(1-\epsilon)|T_0\ra \la T_0| + \epsilon \,|T_1\ra \la T_1|.
\end{equation}
We will assume that the dephasing transformation is applied at the
very first step of the distillation, so $\rho$ has the form~(\ref{diag1}).
Thus the initial state for the elementary distillation subroutine is
\begin{equation}\label{rho_in}
\rho_{in}=\rho^{\otimes 5}=
\sum_{x\in \{0,1\}^5}\! \epsilon^{|x|} (1-\epsilon)^{5-|x|} |T_x\ra\la T_x|,
\end{equation}
where $x=(x_1,\ldots,x_5)$ is a binary string, $|x|$ is the number of 1's in
$x$, and
\[
|T_x\ra\bydef |T_{x_1}\ra\otimes\cdots\otimes |T_{x_5}\ra.
\]

The stabilizers $S_1,\ldots, S_4$ to be measured on the state $\rho_{in}$
correspond to the famous 5-qubit code, see~\cite{BDSW96,LMPZ96}. 
They are defined as follows:
\begin{eqnarray}
\label{5-code}
S_1&=&\sm^x\otimes\sm^z\otimes\sm^z\otimes\sm^x\otimes I, \nn \\
S_2&=&I\otimes\sm^x\otimes\sm^z\otimes\sm^z\otimes\sm^x, \nn \\
S_3&=&\sm^x\otimes I\otimes\sm^x\otimes\sm^z\otimes\sm^z, \nn \\
S_4&=&\sm^z\otimes\sm^x\otimes I\otimes\sm^x\otimes\sm^z.
\end{eqnarray}
This code has a cyclic symmetry, which becomes explicit if we introduce
an auxiliary stabilizer, $S_5=S_1 S_2 S_3 S_4=
\sm^z\otimes\sm^z\otimes\sm^x\otimes I\otimes\sm^x$.
Let $\calL$ be the two-dimensional  code subspace
specified by the conditions $S_j|\Psi\ra=|\Psi\ra$,\, $j=1,\ldots,4$,
and $\Pi$ be the orthogonal projector onto $\calL$:
\begin{equation}
\label{proj}
\Pi = \frac1{16}\prod_{j=1}^4 (I+ S_j).
\end{equation}
It was pointed out in the work~\cite{G97a} that the operators
\[
\hat{X}={(\sm^x)}^{\otimes 5}, \quad
\hat{Y}={(\sm^y)}^{\otimes 5}, \quad
\hat{Z}={(\sm^z)}^{\otimes 5},
\]
and
\begin{equation}\label{hatT}
\hat{T}={(T)}^{\otimes 5}
\end{equation}
commute with $\Pi$, thus preserving the code subspace. Moreover, $\hat{X}$,
$\hat{Y}$, $\hat{Z}$ obey the same algebraic relations as one-qubit Pauli
operators, e.g., $\hat{X}\hat{Y}=i\hat{Z}$. Let us choose a basis in $\calL$
such that $\hat{X}$, $\hat{Y}$, and $\hat{Z}$ become logical Pauli operators
$\sx$, $\sy$, and $\sz$, respectively. How does the operator $\hat{T}$ act in
this basis?  From Eq.~(\ref{T-action}) we immediately get
\[
\hat{T}\hat{X}\hat{T}^\dag = \hat{Z},\quad
\hat{T}\hat{Z}\hat{T}^\dag = \hat{Y},\quad
\hat{T}\hat{Y}\hat{T}^\dag = \hat{X}.
\]
Therefore $\hat{T}$ coincides with the logical operator $T$ up to an overall
phase factor. This factor is fixed by the condition that the logical $T$ has
eigenvalues $e^{\pm i\frac{\pi}{3}}$.

Let us find the eigenvectors of $\hat{T}$ that belong to $\calL$.
Consider two particular states from $\calL$, namely
\[
|T_1^L\ra =\sqrt{6}\, \Pi |T_{00000}\ra,
 \quad  \text{and} \quad
|T_0^L\ra =\sqrt{6}\, \Pi |T_{11111}\ra.
\]
In Appendix~A we show that 
\begin{equation}\label{frac16}
\la T_{00000}|\Pi|T_{00000}\ra = \la T_{11111}|\Pi|T_{11111}\ra = \frac16,
\end{equation}
so that the states $|T_0^L\ra$ and $|T_1^L\ra$ are normalized.
Taking into account that $[\hat{T},\Pi]=0$ and that
\begin{equation}
\label{T5}
\hat{T} |T_x\ra = e^{i\frac{\pi}3 (5-2|x|)}|T_x\ra \quad \text{for all}
\quad x\in \{0,1\}^5,
\end{equation}
we get
\[
\hat{T} |T_1^L\ra = \sqrt{6} \, \hat{T} \Pi |T_{00000}\ra =
\sqrt{6} \, \Pi \hat{T} |T_{00000}\ra = e^{-i\pi/3} |T_1^L\ra.
\]
Analogously, one can check that 
\[
\hat{T} |T_0^L\ra = e^{+i\pi/3} |T_0^L\ra.
\]
It follows that $\hat{T}$ is exactly the logical operator $T$, including the
overall phase, and $|T_0^L\ra$ and $|T_1^L\ra$ are the logical states
$|T_0\ra$ and $|T_1\ra$ (up to some phase factors, which are not important for
us). Therefore we have
\begin{equation}\label{T01-basis}
|T_{0,1}^L\ra\la T_{0,1}^L| =
\Pi \, \frac12 \left[ 
I \pm \frac1{\sqrt{3}} ( \hat{X} + \hat{Y} + \hat{Z} )\right].
\end{equation}

Now we are in a position to describe the syndrome measurement performed on the
state $\rho_{in}$. The unnormalized reduced state corresponding to the trivial
syndrome is as follows:
\begin{equation}
\label{output}
\rho_s=\Pi \rho_{in} \Pi = 
\sum_{x\in \{0,1\}^5}\! \epsilon^{|x|} (1-\epsilon)^{5-|x|} \,
\Pi |T_x\ra\la T_x| \Pi,
\end{equation}
see Eq.~(\ref{rho_in}). The probability for the trivial syndrome
to be observed is 
\[
p_s=\tr\rho_s.
\]
Note that the state $\Pi|T_x\ra$ is an eigenvector of $\hat{T}$ for any $x\in
\{0,1\}^5$. But we know that the restriction of $\hat{T}$ on $\calL$ has
eigenvalues $e^{\pm i \pi/3}$.  At the same time, Eq.~(\ref{T5}) implies that
\[
\hat{T} \Pi |T_x\ra = - \Pi |T_x\ra
\]
whenever $|x|=1$ or $|x|=4$. 
This eigenvalue equation is not a contradiction only if
\[
\Pi |T_x\ra=0\quad \text{for} \quad |x|=1,4.
\]
This equality can be interpreted as an error correction property. Indeed, the
initial state $\rho_{in}$ is a mixture of the desired state $|T_{00000}\ra$
and unwanted states $|T_x\ra$ with $|x|>0$. We can interpret the number of `$1$'
components in $x$ as a number of errors. Once the trivial syndrome has been
measured, we can be sure that either no errors or at least two errors have
occurred. Such error correction, however, is not directly related to the
minimal distance of the code.

It follows from Eq.~(\ref{T5}) that
for $|x|=2,3$ one has $\hat{T}\Pi |T_x\ra = e^{\pm i \pi/3}
\Pi |T_x\ra$, so that $\Pi |T_x\ra$ must be proportional to one of the states
$|T_0^L\ra$, $|T_1^L\ra$. Our observations can be summarized as follows:
\begin{equation}
\label{Pi_map}
\Pi |T_x\ra = \left\{
\ba{rcl}
6^{-1/2} |T_1^L\ra,& {\rm if} & |x| =0,\\[5pt]
0, & {\rm if} & |x|=1, \\[5pt]
a_x |T_0^L\ra, & {\rm if} & |x|=2, \\[5pt]
b_x |T_1^L\ra, & {\rm if} & |x|=3, \\[5pt]
0, & {\rm if} & |x|=4, \\[5pt]
6^{-1/2} |T_0^L\ra,& {\rm if} & |x| =5.
\ea
\right.
\end{equation}
Here the coefficients $a_x$, $b_x$ depend upon $x$ in some way.  The output
state~(\ref{output}) can now be written as
\begin{eqnarray}
\label{output1}
\rho_s &=& 
 \left[ \frac16 \ep^5 + \ep^2 (1-\ep)^3  \sum_{x : |x|=2} |a_x|^2 \right]
|T_0^L\ra\la T_0^L|  \\ 
&& + 
 \left[ \frac16 (1- \ep)^5  + \ep^3 (1-\ep)^2  \sum_{x : |x|=3} |b_x|^2 \right]
|T_1^L\ra\la T_1^L|.\nn
\end{eqnarray}
To exclude the unknown coefficients $a_x$ and $b_x$, we can use the identity 
\[
|T_0^L\ra\la T_0^L| + |T_1^L\ra\la T_1^L| = \Pi = 
\sum_{x\in \{0,1\}^5}\! \Pi |T_x\ra\la T_x| \Pi.
\]
Substituting Eq.~(\ref{Pi_map}) into this identity, we get
\[
\sum_{x : |x|=2} |a_x|^2 = \sum_{x : |x|=3} |b_x|^2 = \frac56.
\]
So the final expression for the output state $\rho_s$ is as follows: 
\begin{eqnarray}
\label{output2}
\rho_s &=& 
\left[\frac{\ep^5 + 5\ep^2 (1-\ep)^3}6\right]|T_0^L\ra\la T_0^L| \nn \\
&&+ 
\left[\frac{(1-\ep)^5 + 5\ep^3(1-\ep)^2}6\right] |T_1^L\ra\la T_1^L|. 
\end{eqnarray}
Accordingly, the probability to observe the trivial syndrome is
\begin{equation}
p_s = \frac{\ep^5 + 5\ep^2 (1-\ep)^3 + 5\ep^3(1-\ep)^2 + (1-\ep)^5}{6}.
\end{equation}

A decoding transformation for the 5-qubit code is a
unitary operator $V\in \Cl{5}$ such that
\[
V \calL = \CC^2\otimes |0,0,0,0\ra.
\]
In other words, $V$ maps the stabilizers $S_j$,\, $j=2,3,4,5$ to $\sz[j]$. The
logical operators $\hat{X}$, $\hat{Y}$, $\hat{Z}$ are mapped to the Pauli
operators $\sx$, $\sy$, $\sz$ acting on the first qubit. From
Eq.~(\ref{T01-basis}) we infer that
\[
V\, |T_{0,1}^L\ra= |T_{0,1}\ra\otimes |0,0,0,0\ra
\]
(maybe up to some phase). The decoding should be followed by an additional
operator $A=\sm^y H \in \Cl{1}$, which swaps the states $|T_0\ra$ and
$|T_1\ra$ (note that for small $\ep$ the state $\rho_s$ is close to
$|T_1^L\ra$, while our goal is to distill $|T_0\ra$).  After that we get a
normalized output state
\[
\rho_{out} = (1-\epsilon_{out})|T_0\ra\la T_0| +
\epsilon_{out} |T_1\ra\la T_1|,
\]
where
\begin{equation}
\label{epsilon}
%\epsilon_{out}= \frac{\ep^5 + 5\ep^2 (1-\ep)^3}
%{\ep^5 + 5\ep^2 (1-\ep)^3 + 5\ep^3(1-\ep)^2 + (1-\ep)^5}.
\epsilon_{out}= \frac{t^5 + 5t^2}{1 + 5t^2 + 5t^3 + t^5 }, \quad\
t=\frac{\epsilon}{1-\epsilon}.
\end{equation}
The plot of the function $\epsilon_{out}(\epsilon)$ is shown on Fig.~2.
%%%%%%%%%%%%%%%%%%%%%%%%%%%%%%%%%%%%%
\begin{figure}
%\psfrag{eps_in}{$\epsilon$}
%\psfrag{eps_out}{$\epsilon_{out}$}
%\psfrag{p_s}{$p_s$}
%\begin{center}
\unitlength=1mm
\begin{picture}(0,0)
\put(37,-50){$\ep$}
\put(37,-100){$\ep$}
\put(-2,-25){$\ep_{out}$}
\put(-2,-75){$p_s$}
\end{picture}
\includegraphics[width=5cm,angle=-90]{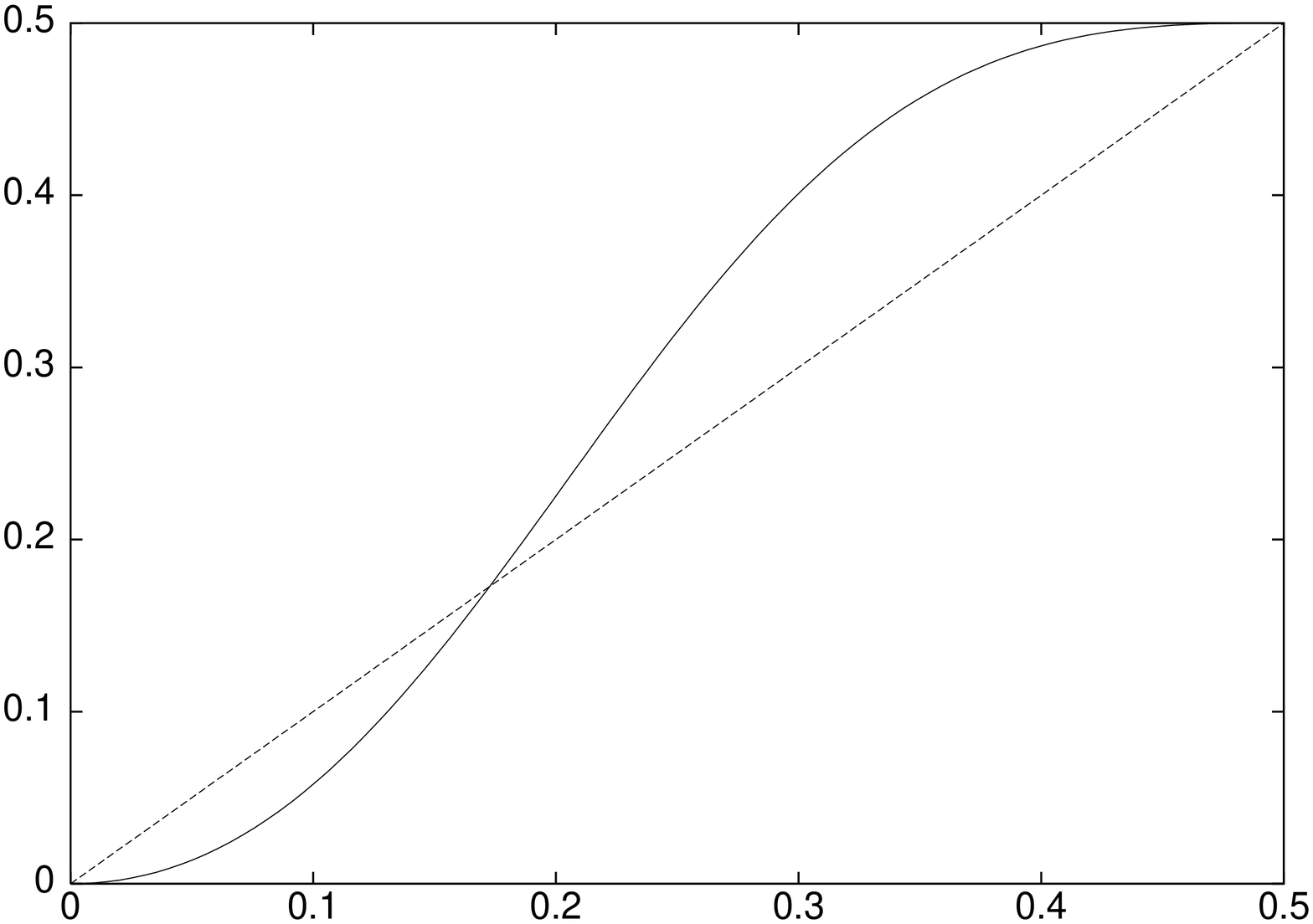}
\includegraphics[width=5cm,angle=-90]{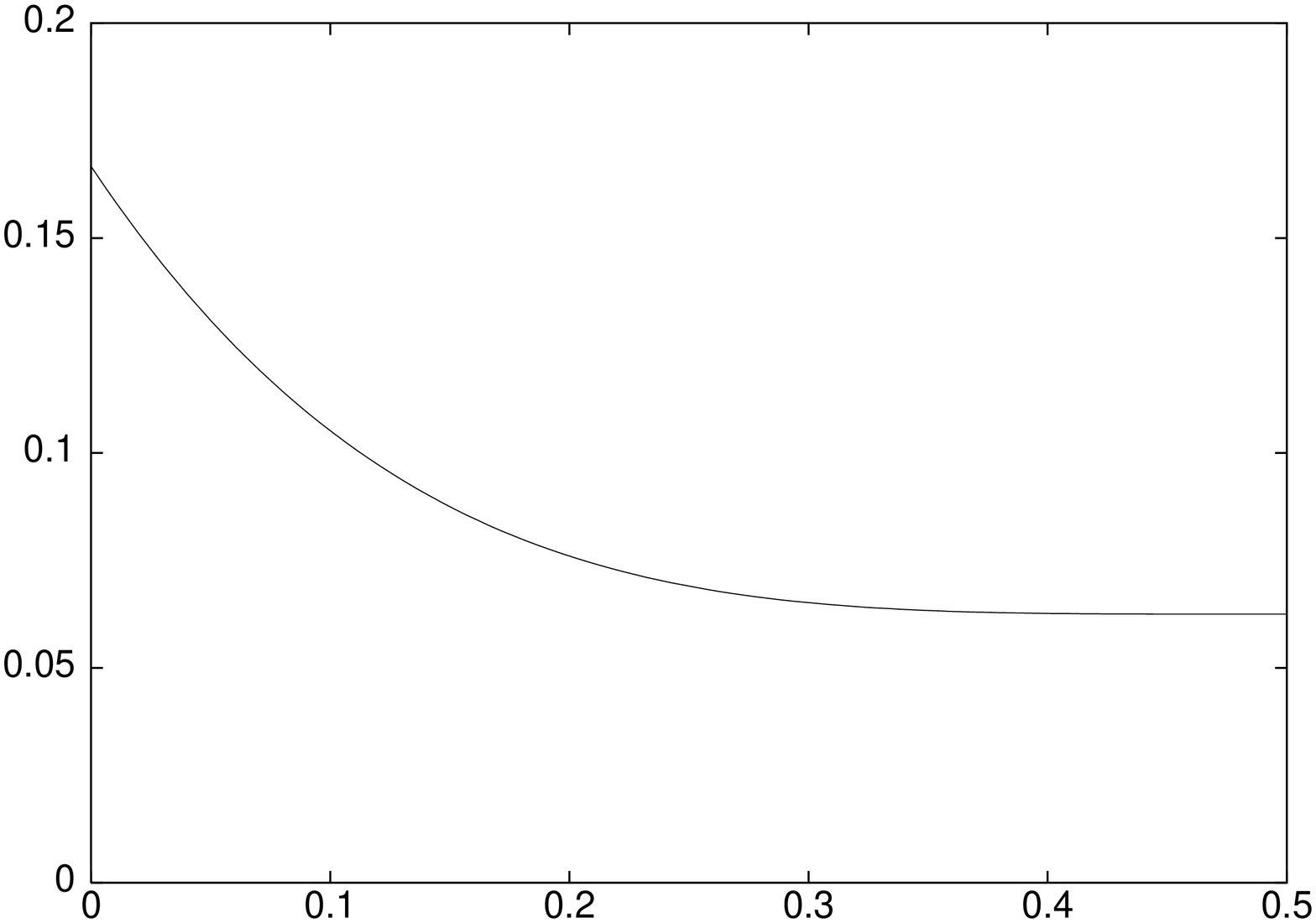}
%\end{center}
\caption{The final error probability $\ep_{out}$ and the probability $p_s$ to
measure the trivial syndrome as functions of the initial error probability
$\ep$ for the \Tt-type states distillation.}
\end{figure}
It indicates that the equation $\epsilon_{out}(\epsilon)=\epsilon$ has only one
non-trivial solution, $\ep=\ep_0\approx 0.173$. The exact value is
\[
\ep_0=\frac12\left( 1 - \sqrt{\frac37} \right).
\]
If $\ep<\ep_0$, we can recursively iterate the elementary distillation
subroutine to produce as good an approximation to the state $|T_0\ra$ as we
wish. On the other hand, if $\ep>\ep_0$, the distillation subroutine increases
the error probability and iterations converge to the maximally mixed
state. Thus $\ep_0$ is a threshold error probability for our scheme. The
corresponding threshold polarization is $1-2\ep_0=\sqrt{3/7}\approx 0.655$.
For a sufficiently small $\ep$, one can use the approximation
$\ep_{out}(\ep)\approx 5\ep^2$.

The probability $p_s=p_s(\ep)$ to measure the trivial syndrome decreases
monotonically from $1/6$ for $\ep=0$ to $1/16$ for $\ep=1/2$, see Fig.~2. In
the asymptotic regime where $\ep$ is small, we can use the approximation
$p_s\approx p_s(0)=1/6$.

Now the construction of the whole distillation scheme is straightforward.  We
start from $n\gg 1$ copies of the state $\rho = (1-\epsilon)|T_0\ra \la T_0| +
\epsilon |T_1\ra\la T_1|$.  Let us split these states into groups containing
five states each and apply the elementary distillation subroutine described
above to each group independently. In some of these groups the distillation
attempt fails, and the outputs of such groups must be discarded. The average
number of ``successful'' groups is obviously $p_s(\ep) (n/5) \approx n/30$ if
$\ep$ is small. Neglecting the fluctuations of this quantity, we can say that
our scheme provides a constant {\it yield} $r=1/30$ of output states that are
characterized by the error probability $\ep_{out}(\ep)\approx 5 \ep^2$.
Therefore we can obtain $r^2 n$ states with $\ep_{out}\approx 5^3 \ep^4$,\,
$r^3 n$ states with $\ep_{out}\approx5^7\ep^{8}$, and so on. We have created a
hierarchy of states with $n$ states on the first level and four or fewer
states on the last level. Let $k$ be the number of levels in this hierarchy
and $\ep_{out}$ the error probability characterizing the states on the last
level.  Up to small fluctuations, the numbers $n$, $k$, $\ep_{out}$ and $\ep$
are related by the following obvious equations:
\begin{equation}
\ep_{out} \approx \frac{1}{5} (5\ep)^{2^k}, \quad r^k n \approx 1.
\end{equation}
Their solution yields Eq.~(\ref{eoutT}).

%%%%%%%%%%%%%%%%%%%%%%%%%%%%%%%%%%%%%%%%%%%%%%%%%%%%%%%%%%%%%%%%%
%%%%%%%%%%%%%%%%%%%%%%%%%%%%%%%%%%%%%%%%%%%%%%%%%%%%%%%%%%%%%%%%%
%%%%%%%%%%%%%%%%%%%%%%%%%%%%%%%%%%%%%%%%%%%%%%%%%%%%%%%%%%%%%%%%%
%%%%%%%%%%%%%%%%%%%%%%%%%%%%%%%%%%%%%%%%%%%%%%%%%%%%%%%%%%%%%%%%%
%%%%%%%%%%%%%%%%%%%%%%%%%%%%%%%%%%%%%%%%%%%%%%%%%%%%%%%%%%%%%%%%%
\section{\label{sec:H} Distillation of \Hh-type magic states}

A distillation scheme for \Hh-type magic states also works by recursive
iteration of a certain elementary distillation subroutine based on a syndrome
measurement for a suitable stabilizer code. Let us start with introducing some
relevant coding theory constructions, which reveal an unusual symmetry of this
code and explain why it is particularly useful for \Hh-type magic states
distillation.

Let $\FF_2^n$ be the $n$-dimensional binary linear space and $A$ be a one-qubit
operator such that $A^{2}=I$. With any binary vector
$u=(u_1,\ldots,u_n)\in\FF_2^n$ we associate the $n$-qubit operator
\[
A(u)=A^{u_1}\otimes A^{u_2}\otimes \cdots\otimes A^{u_n}.
\]
Let $(u,v)=\sum_{i=1}^n u_i v_i \mod{2}$ denote the standard binary inner
product. If $\calL\subseteq \FF_2^n$ is a linear subspace, we denote by
$\calL^\perp$ the set of vectors which are orthogonal to $\calL$. The Hamming
weight of a binary vector $u$ is denoted by $|u|$.  Finally, $u\cdot v \in
\FF_2^n$ designates the bitwise product of $u$ and $v$ i.e.,  $(u\cdot v)_i
=u_i v_i$.

A systematic way of constructing stabilizer codes was suggested by Calderbank,
Shor, and Steane, see~\cite{CSS96,CSS96'}. Codes that can be described in this
way will be referred to as \emph{standard CSS codes}. In addition, we consider
their images under an arbitrary unitary transformation $V\in U(2)$ applied to
every qubit. Such ``rotated'' codes will be called \emph{CSS codes}.
\begin{dfn}
Consider a pair of one-qubit Hermitian operators
$A, B$ such that
\[
A^2=B^2=I, \quad AB=-BA,
\]
and a pair of binary vector spaces
$\calL_A, \calL_B \subseteq \FF_2^n$, such that
\[
(u,v)=0 \quad \text{for all}\,\ u\in \calL_A,\ v\in \calL_B.
\]
A quantum code $\CSS(A,\calL_A;\,B,\calL_B)$ is a decomposition
\begin{equation}\label{syndrome-decomposition}
(\CC^2)^{\otimes n}=\bigoplus_{\mu\in \calL_A^*}
\bigoplus_{\eta\in \calL_B^*}
\calH(\mu,\eta),
\end{equation}
where the subspace $\calH(\mu,\eta)$ is defined by the conditions
\[
A(u)|\Psi\ra=(-1)^{\mu(u)}|\Psi\ra,\quad
B(v)|\Psi\ra=(-1)^{\eta(v)}|\Psi\ra
\]
for all $u\in \calL_A$ and $v\in \calL_B$. 
The linear functionals $\mu$ and $\eta$ are refereed to as
A-syndrome and B-syndrome, respectively. The subspace $\calH(0,0)$
corresponding to the trivial syndromes $\mu=\eta=0$
is called the code subspace.
\end{dfn}
The subspaces $\calH(\mu,\eta)$ are well defined since 
the operators $A(u)$ and $B(v)$ commute for any $u\in \calL_A$
and $v \in \calL_B$:
\[
A(u)B(v)=(-1)^{(u,v)}B(v)A(u)=B(v)A(u).
\]
The number of logical qubits in a CSS code is
\[
k\bydef\log_2\bigl(\dim{\calH(0,0)}\bigr)=n-\dim{\calL_A} - \dim{\calL_B}.
\]
Logical operators preserving the subspaces $\calH(\mu,\eta)$ can be
chosen as
\[
\left\{ A(u):\ u\in \left.\calL_B^\perp\right/ \calL_A \right\} 
\quad \text{and} \quad
\left\{ B(v):\ v\in \left.\calL_A^\perp\right/ \calL_B \right\}.
\]
(By definition, $\calL_A\subseteq \calL_B^\perp$ and $\calL_B\subseteq
\calL_A^\perp$, so the factor spaces are well defined.)  In the case where $A$
and $B$ are Pauli operators, we get a standard CSS code. Generally, $A=V\sz
V^\dag$ and $B=V\sx V^\dag$ for some unitary operator $V\in SU(2)$, so an
arbitrary CSS code can be mapped to a standard one by a suitable bitwise
rotation. By a syndrome measurement for a CSS code we mean a projective
measurement associated with the decomposition~(\ref{syndrome-decomposition}).

Consider a CSS code such that some of the operators $A(u)$, $B(v)$ do not
belong to the Pauli group $P(n)$.  Let us pose this question: can one perform
a syndrome measurement for this code by operations from $\calO_{ideal}$ only?
It may seem that the answer is `no', because by definition of $\calO_{ideal}$
one cannot measure an eigenvalue of an operator unless it belongs to the Pauli
group. Surprisingly, this naive answer is wrong.  Indeed, imagine that we have
measured part of the operators $A(u)$, $B(v)$ (namely, those ones that belong
to the Pauli group). Now we may restrict the remaining operators to the
subspace corresponding to the obtained measurement outcomes. It may happen
that the restriction of some unmeasured operator, $A(u)$, which does not
belong to the Pauli group, coincides with the restriction of some other
operator $\tilde{A}(\tilde{u})\in P(n)$.  If this is the case, we can safely
measure $\tilde{A}(\tilde{u})$ instead of $A(u)$. The 15-qubit code that we
use for the distillation is actually the simplest (to our knowledge) CSS code
exhibiting this strange behavior. We now come to an explicit description of
this code.

Consider a function $f$ of four Boolean variables. Denote by $[f]\in
\FF_2^{15}$ the table of all values of $f$ except $f(0000)$. The table is
considered as a binary vector i.e.,
\[
[f]=\bigl( f(0001),\, f(0010),\, f(0011), \,\ldots\, ,f(1111)\bigr).
\]
Let $\calL_1$ be the set of all vectors $[f]$, where $f$ is a linear function
satisfying $f(0)=0$. In other words, $\calL_1$ is the linear subspace spanned
by the four vectors $[x_j]$,\, $j=1,2,3,4$ (where $x_{j}$ is an
indicator function for the $j$th input bit):
\[
\calL_1=\text{lin.span}\bigl([x_1], [x_2], [x_3], [x_4]\bigr).
\]
Let also $\calL_2$ be the set of all vectors $[f]$, where $f$ is a polynomial
of degree at most $2$ satisfying $f(0)=0$.  In other words, $\calL_2$ is the
linear subspace spanned by the four vectors $[x_j]$ and the six vectors $[x_i
x_j]$:
\begin{equation}
\calL_2 = \text{lin.span} \left(
\ba{c}
[x_1], [x_2], [x_3], [x_4], [x_1 x_2], [x_1 x_3], \\
\, [x_1 x_4], [x_2 x_3], [x_2 x_4], [x_3 x_4]
\ea
\right).
\end{equation}
The definition of $\calL_1$ and $\calL_2$ resembles the definition of
punctured Reed-Muller codes of order one and two, respectively,
see~\cite{MacWilliams}. Note also that $\calL_1$ is the dual space for the
15-bit Hamming code.

The relevant properties of the subspaces $\calL_j$ are
stated in the following lemma.
\begin{lemma}\label{lemma:weights}
{\mbox{}}\\
1) For any $u \in \calL_1$ one
has $|u|\equiv 0\pmod8$.\\
2) For any  $v\in \calL_2$ one has $|v|\equiv 0\pmod2$.\\
3) Let $[1]$ be the unit vector $(1,1,\ldots,1, 1)$. Then\\ \hbox to 3mm{}
$\calL_1^\perp = \calL_2 \oplus [1]$ and $\calL_2^\perp
= \calL_1 \oplus [1]$.\\
4) For any vectors $u,v \in \calL_1$ one has
$|u\cdot v|\equiv 0\pmod4$.\\
5) For any vectors $u\in\calL_1$ and $v\in \calL_2^\perp$ one has\\
\hbox to 3mm{} 
$|u\cdot v|\equiv 0\pmod4$.
\end{lemma}
\begin{proof} \mbox{} \\
%%%
1) Any linear function $f$ on $\FF_2^4$ satisfying $f(0)=0$ takes value $1$
exactly eight times (if $f\ne 0$) or zero times (if $f=0$).\\
%%%
2) All basis vectors of $\calL_2$ have weight equal to $8$ (the vectors
$[x_i]$) or $4$ (the vectors $[x_i x_j]$).  By linearity, all elements of
$\calL_2$ have even weight. \\
%%%
3) One can easily check that all basis vectors of $\calL_1$ are orthogonal to
all basis vectors of $\calL_2$, therefore $\calL_1\subseteq \calL_2^\perp$,\,
$\calL_2\subseteq \calL_1^\perp$.  Besides, we have already proved that
$[1]\in \calL_1^\perp$ and $[1]\in \calL_2^\perp$. Now the statement follows
from dimension counting, since $\dim{\calL_1}=4$ and $\dim{\calL_2}=10$.\\
%%%
4) Without loss of generality we may assume that $u\not=0$ and $v\not=0$. If
$u=v$, the statement has been already proved, see property~1. If $u \ne v$,
then $u=[f]$, $v=[g]$ for some linearly independent linear functions $f$ and
$g$. We can introduce new coordinates $(y_1,y_2,y_3,y_4)$ on $\FF_2^4$ such
that $y_1=f(x)$ and $y_2=g(x)$.  Now $|u\cdot v|=\bigl|[y_1 y_2]\bigr| =4$.\\
%%%
5) Let $u\in \calL_1$ and $v\in \calL_2^\perp$. Since $\calL_2^\perp =\calL_1
\oplus [1]$, there are two possibilities: $v\in\calL_1$ and $v=[1]+w$ for some
$w\in \calL_1$. The first case has been already considered. In the second case
we have
\[
|u\cdot v|=\sum_{j=1}^{15} u_j(1-w_j)= |u| - |u\cdot w|.
\]
It follows from properties~1 and~4 that $|u\cdot v|\equiv 0\pmod4$. 
\end{proof}

Now consider the one-qubit Hermitian operator
\[
A=\frac1{\sqrt{2}} (\sx+\sy) = \left( \ba{cc} 0 & e^{-i\frac{\pi}4} \\
e^{+i\frac{\pi}4} & 0 \\ \ea \right)=
e^{-i\frac{\pi}4} K\sx,
\]
where $K$ is the phase shift gate, see Eq.~(\ref{gates}). By definition, $A$
belongs to the Clifford group $\Cl{1}$.  One can easily check that $A^2=I$ and
$A\sz = -\sz A$, so the code $\CSS(\sz, \calL_2;\, A, \calL_1)$ is well
defined.  We claim that its code subspace coincides with the code subspace of
a certain stabilizer code.
\begin{lemma}\label{lemma:CSS=CSS}
Consider the decomposition
\[
(\CC^2)^{\otimes 15}=\bigoplus_{\mu\in \calL_2^*}\bigoplus_{\eta\in \calL_1^*}
\calH(\mu,\eta),
\]
associated with the code $\CSS(\sz,\calL_2; \, A, \calL_1)$ and the
decomposition
\[
(\CC^2)^{\otimes 15}=\bigoplus_{\mu\in \calL_2^*}\bigoplus_{\eta\in \calL_1^*}
\calG(\mu,\eta),
\]
associated with the stabilizer code $\CSS(\sz,\calL_2; \, \sx, \calL_1)$. For
any syndrome $\eta\in \calL_1^*$ one has
\[
\calH(0,\eta)=\calG(0,\eta).
\]
Moreover, for any $\mu\in\calL_2^*$ there exists some $w\in\FF_{2}^{15}$ such
that for any $\eta\in \calL_1^*$
\begin{equation}\label{rotated_code}
\calH(\mu,\eta)=A(w)\calG(0,\eta).
\end{equation}
\end{lemma}
This Lemma provides a strategy to measure a syndrome of the code
$\CSS(\sz,\calL_2; \, A, \calL_1)$ by operations from
$\calO_{ideal}$. Specifically, we measure $\mu$ (i.e., the $\sigma^{z}$ part
of the syndrome) first, compute $w=w(\mu)$, apply $A(w)^{\dag}$, measure
$\eta$ using the stabilizers $\sigma^{x}([x_{j}])$, and apply $A(w)$.

\begin{proof}[Proof of the Lemma]
Consider an auxiliary subspace,
\[
\calH=\bigoplus_{\eta\in \calL_1^*} \calH(0,\eta)=
\bigoplus_{\eta\in\calL_1^*} \calG(0,\eta)
\]
corresponding to the trivial $\sz$-syndrome for both CSS codes. Each
state $|\Psi\ra\in \calH(0)$ can be represented as
\[
|\Psi\ra = \sum_{v\in \calL_2^\perp} c_v |v\ra,
\]
where $c_v$ are some complex amplitudes and $|v\ra=|v_1,\ldots,v_{15}\ra$ are
vectors of the standard basis. Let us show that
\[
A(u)|\Psi\ra = \sx(u)|\Psi\ra\quad \text{for any}\ 
|\Psi\ra\in \calH,\,\ u\in\calL_1.
\]
To this end, we represent $A$ as $\sx e^{i\pi/4} K^\dag$. For any $u\in
\calL_1$ and $v\in\calL_2^{\perp}$ we have
\[
A(u)|v\ra \,=\,
\sigma^{x}(u)\, e^{i\frac{\pi}{4}|u|-i\frac{\pi}{2} |u\cdot v|}|v\ra
\,=\, \sigma^{x}(u) |v\ra,
\]
because $|u|\equiv 0\pmod{8}$ and $|u\cdot v|\equiv 0\pmod{4}$ (see
Lemma~\ref{lemma:weights}, parts~1 and~5).

Since for any $u\in \calL_1$ the operators $A(u)$ and $\sx(u)$ act on $\calH$
in the same way, their eigenspaces must coincide i.e.,
$\calH(0,\eta)=\calG(0,\eta)$ for any $\eta\in\calL_1^*$.

Let us now consider the subspace $\calH(\mu,\eta)$ for arbitrary
$\mu\in\calL_2^*$,\, $\eta\in\calL_1^*$. By definition, $\mu$ is a linear
functional on $\calL_2\subseteq\FF_{2}^{15}$; we can extend it to a linear
functional on $\FF_{2}^{15}$ i.e., represent it in the form $\mu(v)=(w,v)$
for some $w\in\FF_{2}^{15}$. Then for any $|\Psi\rangle\in\calH(\mu,\eta)$,\,
$v\in\calL_2$, and $u\in\calL_1$ we have
\begin{align*}
\sigma^{z}(v)A(w)^{\dag}|\Psi\rangle &=
(-1)^{(w,v)}A(w)^{\dag}\sigma^{z}(v)|\Psi\rangle
=A(w)^{\dag}|\Psi\rangle,\\[3pt]
A(u)A(w)^{\dag}|\Psi\rangle &=
A(w)^{\dag}A(u)|\Psi\rangle=(-1)^{\eta(v)}A(w)^{\dag}|\Psi\rangle
\end{align*}
(as $\sigma^{z}$ and $A$ anticommute), hence
$A(w)^{\dag}|\Psi\rangle\in\calH(0,\eta)$. Thus,
\[
\calH(\mu,\eta)=A(w)\calH(0,\eta)=A(w)\calG(0,\eta).
\]
\end{proof}

Lemma~\ref{lemma:CSS=CSS} is closely related to an interesting property of the
stabilizer code $\CSS(\sz,\calL_2; \,\sx,\calL_1)$, namely the existence of a
non-Clifford automorphism~\cite{KLZ96}.  Consider a one-qubit unitary operator
$W$ such that
\[
W \sz W^\dag =\sz \quad \text{and} \quad
W \sx W^\dag = A.
\]
It is defined up to an overall phase and obviously does not belong to the
Clifford group $\Cl{1}$. However, the bitwise application of $W$ i.e., the
operator $W^{\otimes 15}$ preserves the code subspace $\calG(0,0)$. Indeed,
$W^{\otimes 15}\calG(0,0)$ corresponds to the trivial syndrome of the code
\[
\CSS(W\sz W^{\dag},\calL_2;\,W\sx W^{\dag},\calL_1) = 
\CSS(\sz,\calL_2;\,A,\calL_1).
\]
Thus $W^{\otimes 15}\calG(0,0)=\calH(0,0)$. But $\calH(0,0)=\calG(0,0)$ due to
the lemma.

Now we are in a position to describe the distillation scheme and to estimate
its threshold and yield. Suppose we are given 15 copies of the state $\rho$,
and our goal is to distill one copy of an \Hh-type magic state.  We will
actually distill the state
\[
|A_0\ra \bydef \frac1{\sqrt{2}} \left( |0\ra + e^{i\frac{\pi}4} |1\ra \right)=
e^{i\frac{\pi}8} H K^\dag |H\ra.
\]
Note that $|A_0\ra$ is an eigenstate of the operator $A$; specifically,
$A|A_0\ra=|A_0\ra$. Let us also introduce the state 
\[
|A_1\ra=\sz |A_0\ra,
\]
which satisfies $A|A_1\ra=-|A_1\ra$. 
Since the Clifford group $\Cl{1}$ acts transitively on the set of \Hh-type
magic states, we can assume that the fidelity
between $\rho$ and $|A_0\ra$ is the maximum one among all \Hh-type magic states,
so that
\[
F_H(\rho) = \sqrt{\la A_0|\rho|A_0\ra}.
\]
As in Sec.~\ref{sec:T} we define the initial error probability
\[
\ep \bydef 1 - [F_H(\rho)]^2 = \la A_1|\rho|A_1\ra.
\]
Applying the dephasing transformation
\[
D(\eta)=\frac12 \left( \eta + A\eta A^\dag\right)
\]
to each copy of $\rho$, we can guarantee that $\rho$ is diagonal in the $\{
A_0, A_1\}$ basis i.e.,
\[
\rho = D(\rho)= (1-\ep) |A_0\ra\la A_0| + \ep |A_1\ra\la A_1|.
\]
Since $A\in \Cl{1}$, the dephasing transformation  can be realized by
operations from $\calO_{ideal}$. Thus out initial state is
\begin{equation}\label{H-initial}
\rho_{in} = \rho^{\otimes 15}
=\sum_{u\in \FF_2^{15}} \ep^{|u|} (1-\ep)^{15-|u|}\,
|A_u\ra\la A_u|,
\end{equation}
where $|A_u\ra\bydef |A_{u_{0}}\ra\otimes\cdots\otimes|A_{u_{15}}\ra$.

According to the remark following
the formulation of Lemma~\ref{lemma:CSS=CSS}, we can
measure the syndrome $(\mu,\eta)$ of the code $\CSS(\sz, \calL_2;\, A,
\calL_1)$ by operations from $\calO_{ideal}$ only. Let us follow this scheme,
omitting the very last step. So, we begin with the state $\rho_{in}$, measure
$\mu$, compute $w=w(\mu)$, apply $A(w)^{\dag}$, and measure $\eta$. We
consider the distillation attempt successful if $\eta=0$. The measured value of
$\mu$ is not important at this stage. In fact, for any $\mu\in\calL_2^*$ the
unnormalized post-measurement state is
\[
\rho_{s}=\Pi A(w)^{\dag}\rho_{in} A(w)\Pi = \Pi\rho_{in}\Pi.
\]
In this equation
$\Pi$ is the projector onto the code subspace $\calH(0,0)=\calG(0,0)$
i.e., $\Pi=\Pi_{z}\Pi_{A}$ for
\begin{equation}\label{AZ-proj}
\Pi_{z}=\frac1{|\calL_2|} \sum_{v\in \calL_2} \sz(v),\quad\
\Pi_{A}=\frac1{|\calL_1|} \sum_{u\in \calL_1} A(u).
\end{equation}

Let us compute the state $\rho_{s}=\Pi\rho_{in}\Pi$. Since
\[
A(u)|A_w\ra=(-1)^{(u,w)} |A_w\ra, \qquad \sz(v) |A_w\ra = |A_{w+v}\ra,
\]
one can easily see that $\Pi_{A}|A_w\ra=|A_w\ra$ if $w\in\calL_1^\perp$,
otherwise $\Pi_{A}|A_w\ra=0$. On the other hand, $\Pi_{z}|A_w\ra$ does not
vanish and depends only on the coset of $\calL_{2}$ that contains $w$. There
are only two such cosets in $\calL_1^\perp$ (because
$\calL_1^\perp=\calL_2\oplus [1]$, see Lemma~\ref{lemma:weights}), and the
corresponding projected states are:
\begin{eqnarray}\label{A-logical}
|A_0^L\ra &\bydef& \sqrt{|\calL_2|}\,\Pi_{z}|A_{0\ldots 0}\rangle =
\frac1{\sqrt{|\calL_2|}} \sum_{v\in \calL_2} |A_v\ra, \\
|A_1^L\ra &\bydef& \sqrt{|\calL_2|}\,\Pi_{z}|A_{1\ldots 1}\rangle =
\frac1{\sqrt{|\calL_2|}} \sum_{v\in \calL_2} |A_{v+[1]}\ra. \nn
\end{eqnarray}
The states $|A_{0,1}^L\ra$ form an orthonormal basis of the code subspace.
The projections of $|A_w\ra$ for $w\in \calL_1^\perp$ onto the code subspace
are given by these formulas:
\begin{eqnarray}
\Pi |A_w\ra = \frac1{\sqrt{|\calL_2|}} |A_0^L\ra &\text{ if }& 
w\in \calL_2, \nn\\
\Pi |A_w\ra = \frac1{\sqrt{|\calL_2|}} |A_1^L\ra &\text{ if }&
w\in \calL_2 + [1]. \nn
\end{eqnarray}
Now the unnormalized final state $\rho_{s}=\Pi\rho_{in}\Pi$ can be expanded as
\begin{eqnarray}
\rho_{s} &=& 
\frac1{|\calL_2|}\sum_{v\in \calL_2} (1-\ep)^{15-|v|}\ep^{|v|}\, 
|A_0^L\ra\la A_0^L| \nn\\
&& {}+
\frac1{|\calL_2|}\sum_{v\in \calL_2} \ep^{15-|v|}(1-\ep)^{|v|}\,
|A_1^L\ra\la A_1^L|. \nn
\end{eqnarray}
The distillation succeeds with probability
\[
p_{s}\,=\,|\calL_{2}|\tr\rho_{s}\,=\,
\sum_{v\in \calL_1^\perp} \ep^{15-|v|}(1-\ep)^{|v|}.
\]
(The factor $|\calL_{2}|$ reflects the number of possible values of $\mu$,
which all give rise to the same state $\rho_{s}$.)

To complete the distillation procedure, we need to apply a decoding
transformation that would map the two-dimensional subspace
$\calH(0,0)\subset(\CC^{2})^{\otimes 15}$ onto the Hilbert space of one
qubit. Recall that $\calH(0,0)=\calG(0,0)$ is the code subspace of the
stabilizer code $\CSS(\sz,\calL_2;\,\sx,\calL_1)$.  Its logical Pauli
operators can be chosen as
\[
\hat{X}={(\sx)}^{\otimes 15}, \quad \hat{Y}={(\sy)}^{\otimes 15}, \quad
\hat{Z}=-{(\sz)}^{\otimes 15}.
\]
It is easy to see that $\hat{X}$, $\hat{Y}$, $\hat{Z}$ obey the correct
algebraic relations and preserve the code subspace. The decoding can be
realized as a Clifford operator $V\in\Cl{15}$ that maps $\hat{X}$, $\hat{Y}$,
$\hat{Z}$ to the Pauli operators $\sx$, $\sy$, $\sz$ acting on the first
qubit. (The remaining fourteen qubits become unentangled with the first one,
so we can safely disregard them.) Let us show that the logical state
$|A_{0}^L\ra$ is transformed into $|A_{0}\ra$ (up to some phase). For this, it
suffices to check that $\la A_{0}^L|\hat{X}|A_{0}^L\ra=\la
A_{0}|\sigma^{x}|A_{0}\ra$,\, $\la A_{0}^L|\hat{Y}|A_{0}^L\ra=\la
A_{0}|\sigma^{y}|A_{0}\ra$, and $\la A_{0}^L|\hat{Z}|A_{0}^L\ra=\la
A_{0}|\sigma^{z}|A_{0}\ra$. Verifying these identities becomes a
straightforward task if we represent $|A_{0}^L\ra$ in the standard basis:
\begin{align*}
|A_{0}^L\ra\, &=\,|\calL_{2}|^{1/2} 2^{-15/2}\sum_{u\in\calL_{2}^{\perp}}
e^{i\frac{\pi}{4}|u|} |u\rangle\\
&=\, 2^{-5/2}\sum_{u\in\calL_1}
\Bigl(\bigl|u\bigr\rangle+e^{-i\frac{\pi}{4}}\bigl|u+[1]\bigr\rangle\Bigr).
\end{align*}

To summarize, the distillation subroutine consists of the following steps.
\begin{enumerate}
\item Measure eigenvalues of the Pauli operators $\sz([x_j])$, $\sz([x_jx_k])$ (for
$j,k=1,2,3,4$).  The outcomes determine the $\sz$-syndrome, $\mu\in\calL_2^*$.
\item Find $w=w(\mu)\in\FF_2^{15}$ such that $(w,v)=\mu(v)$ for any
$v\in\calL_{2}$.
\item Apply the correcting operator $A(w)^{\dagger}$.
\item Measure eigenvalues of the operators $\sx([x_j])$.  The outcomes
determine the $A$-syndrome, $\eta\in\calL_1^*$.
\item Declare failure if $\eta\ne 0$, otherwise proceed to the next
step.
\item Apply the decoding transformation, which takes the code subspace to the
Hilbert space of one qubit.
\end{enumerate}
The subroutine succeeds with probability
\begin{equation}\label{probsuccess}
p_{s}=\sum_{v\in \calL_1^\perp} \ep^{15-|v|}(1-\ep)^{|v|}.
\end{equation}
In the case of success, it produces the normalized output state
\begin{equation}\label{rho_out}
\rho_{out} = (1-\ep_{out})|A_0\ra\la A_0| + \ep_{out}|A_1\ra\la A_1|
\end{equation}
characterized by the error probability
\begin{equation}\label{errprob}
\ep_{out}=p_{s}^{-1}\sum_{v\in \calL_2} \ep^{15-|v|}(1-\ep)^{|v|}.
\end{equation}

The sums in Eqs.~(\ref{probsuccess}) and~(\ref{errprob}) are special forms of
so-called weight enumerators. The {\it weight enumerator} of a subspace
$\calL\subseteq \FF_2^n$ is a homogeneous polynomial of degree $n$ in two
variables, namely
\[
W_{\calL}(x,y)=\sum_{u\in \calL} x^{n-|u|} y^{|u|}.
\]
In this notation,
\[
p_{s}=W_{\calL_1^\perp}(\ep,1-\ep),\quad\
\ep_{out}= \frac{W_{\calL_2}(\ep,1-\ep)}{W_{\calL_1^\perp}(\ep,1-\ep)}.
\]
The MacWilliams identity~\cite{MacWilliams} relates the weight enumerator of
$\calL$ to that of $\calL^\perp$:
\[
W_{\calL}(x,y) = \frac1{|\calL^\perp|} W_{\calL^\perp}(x+y,x-y).
\]
Applying this identity and taking into account that $\calL_2^\perp =
\calL_1\oplus[1]$ and that $|u|\equiv 0\pmod2$ for any $u\in\calL_1$ (see
Lemma~\ref{lemma:weights}), we get:
\begin{equation}\label{ep_out1}
p_{s}=\frac{1}{16}W_{\calL_1}(1,1-2\ep),\quad
\ep_{out}=\frac12\left(
1- \frac{W_{\calL_1}(1-2\ep,1)}{W_{\calL_1}(1,1-2\ep)}\right). 
\end{equation}
The weight enumerator of the subspace $\calL_1$ is particularly simple:
\[
W_{\calL_1}(x,y) = x^{15} + 15 x^7 y^8.
\]
Substituting this expression into Eq.~(\ref{ep_out1}), we arrive at the
following formulas:
\begin{equation}\label{p_s}
p_{s}=\frac{1+15(1-2\ep)^{8}}{16},
\end{equation}
\begin{equation}\label{ep_out}
\ep_{out} = 
\frac{1- 15 (1-2\ep)^7 + 15 (1-2\ep)^8 - (1-2\ep)^{15}}
{2\bigl(1 + 15 (1-2\ep)^8\bigr)}.
\end{equation}
The function $\ep_{out}(\ep)$ is plotted on Fig.~3. Solving the equation
$\ep_{out}(\ep)=\ep$ numerically, we find the threshold error probability:
\begin{equation}
\ep_0\approx 0.141.
\end{equation}
\begin{figure}
\unitlength=1mm
\begin{picture}(0,0)
\put(37,-50){$\ep$}
\put(-2,-25){$\ep_{out}$}
\end{picture}
\includegraphics[width=5cm,angle=-90]{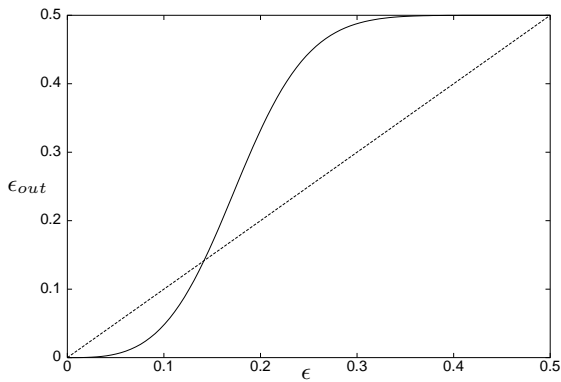}
%\end{center}
\caption{The final error probability $\ep_{out}(\ep)$
for the \Hh-type states distillation.}
\end{figure}

Let us examine the asymptotic properties of this scheme.  For small $\ep$ the
distillation subroutine succeeds with probability close to $1$, therefore the
yield is close to $1/15$. The output error probability is
\begin{equation}\label{35ep^3}
\ep_{out}\approx 35 \ep^3.
\end{equation}
Now suppose that the subroutine is applied recursively. From $n$ copies of the
state $\rho$ with a given $\ep$, we distill one copy of the magic state
$|A_0\ra$ with the final error probability
\[
\ep_{out}(n,\ep)\approx\frac1{\sqrt{35}}
\left( \sqrt{35}\ep\right)^{\displaystyle 3^k},\quad
15^k\approx n,
\]
where $k$ is the number of recursion levels (here we neglect the fluctuations
in the number of successful distillation attempts).  Solving these equation,
we obtain the relation
\begin{equation}\label{H-eff}
\ep_{out}(n,\ep)\sim \left( \sqrt{35} \ep\right)^{\displaystyle n^\xi},
\quad\ \xi=1/\log_{3}15\approx 0.4.
\end{equation}
It characterizes the efficiency of the distillation scheme.

\section{\label{sec:conclusion} Conclusion and some open problems}

We have studied a simplified model of fault-tolerant quantum computation in
which operations from the Clifford group are realized exactly, whereas
decoherence occurs only during the preparation of nontrivial ancillary
states. The model is fully characterized by a one-qubit density matrix $\rho$
describing these states. It is shown that a good strategy for simulating
universal quantum computation in this model is ``magic states
distillation''. By constructing two particular distillation schemes we find a
threshold polarization of $\rho$ above which the simulation is possible.

The most exciting open problem is to understand the computational power of the
model in the region of parameters $1<|\rho_x|+|\rho_y|+|\rho_z|\le 3/\sqrt{7}$
(which corresponds to $F_T^* < F_T(\rho)\le F_T$, see
section~\ref{sec:intro}). In this region, the distillation scheme based on the
5-quit code does not work, while the Gottesman-Knill theorem does not yet
allow the classical simulation. One possibility is that a transition from
classical to universal quantum behavior occurs on the octahedron boundary,
$|\rho_x|+|\rho_y|+|\rho_z|=1$.

To prove the existence of such a transition, one it suffices to construct a
\Tt-type states distillation scheme having the threshold fidelity $F_T^*$. A
systematic way of constructing such schemes is to replace the 5-qubit by a
$GF(4)$-linear stabilizer code.  A nice property of these codes is that the
bitwise application of the operator $T$ preserves the code subspace and acts
on the encoded qubit as $T$, see~\cite{CRSS96} for more details.  One can
check that the error-correcting effect described in Sec.~\ref{sec:T} takes
place for an arbitrary $GF(4)$-linear stabilizer code, provided that the
number of qubits is $n=6k-1$ for any integer $k$.  Unfortunately, numerical
simulations we performed for some codes with $n=11$ and $n=17$ indicate that the
threshold fidelity increases as the number of qubits increases.  So it may
well be the case that the 5-qubit code is the best $GF(4)$-linear code as far
as the distillation is concerned.

From the experimental point of view, an exciting open problem is to design a
physical system in which reliable storage of quantum information and its
processing by Clifford group operations is possible. Since our simulation
scheme tolerates strong decoherence on the ancilla preparation stage, such a
system would be a good candidate for a practical quantum computer.

\begin{acknowledgments}
We thank Mikhail Vyalyi for bringing to our attention many useful facts about
the Clifford group. This work has been supported in part by the National
Science Foundation under Grant No.\ EIA-0086038.
\end{acknowledgments}

\appendix
\section{}
The purpose of this section is to prove Eq.~(\ref{frac16}).
Let us introduce this notation:
\[
|\hat{T}_0\ra=|T_{00000}\ra\quad \text{and} \quad
|\hat{T}_1\ra=|T_{11111}\ra.
\]
Consider the set $S_+(5)\subset S(5)$ consisting of all possible tensor
products of the Pauli operators $\sx$, $\sy$, $\sz$ on five qubits (clearly,
$|S_+(5)|=4^5=|S(5)|/2$ since elements of $S(5)$ may have a plus or minus
sign). For each $g\in S_+(5)$ let $|g|\in [0,5]$ be the number of qubits on
which $g$ acts nontrivially (e.g. $|\sx\otimes\sx\otimes\sy\otimes I\otimes
I|=3$).  We have
\[
|\hat{T}_{0}\ra\la \hat{T}_{0}|=\frac1{2^5}
\sum_{g\in S_+(5)} \left(\frac1{\sqrt{3}}\right)^{|g|} g.
\]
Now let us expand the formula~(\ref{proj}) for the projector $\Pi$.  Denote by
$G\subset P(5)$ the Abelian group generated by the stabilizers $S_1, S_2, S_3,
S_4$.  It consists of sixteen elements.  Repeatedly conjugating the stabilizer $S_1$ by
the operator $\hat{T}=T^{\otimes 5}$, we get three elements of $G$:
\begin{eqnarray}
S_1 &=& \sx\otimes\sz\otimes\sz\otimes\sx\otimes I,\nn \\
S_1 S_3 S_4 &=& \sz\otimes\sy\otimes\sy\otimes\sz\otimes I,\nn \\
S_3 S_4 &=& \sy\otimes\sx\otimes\sx\otimes\sy\otimes I. \nn
\end{eqnarray}
Due to the cyclic symmetry mentioned in section~\ref{sec:T}, the 15 cyclic
permutations of these elements also belong to $G$; together with the identity
operator they exhaust the group $G$. Thus $G\subset S_+(5)$, and we have
\[
\Pi = \frac1{16} \sum_{h\in G} h.
\]
Taking into account that $\tr(gh)=2^5 \delta_{g,h}$ for any
$g,h\in S_+(5)$, we get
\begin{eqnarray}
\la\hat{T}_{0}|\Pi|\hat{T}_{0}\ra &=&
\frac1{2^9}\sum_{h\in G} \sum_{g\in S_+(5)} 
3^{-|g|/2} \tr{(gh)} \nn \\
&& {}
= \frac1{16} \sum_{g\in G} 3^{-|g|/2} =
\frac16.\nn
\end{eqnarray}
Similar calculations show that
$\la\hat{T}_{1}|\Pi|\hat{T}_{1}\ra=\frac{1}{6}$.

\end{document}